\documentclass[runningheads]{llncs}

\usepackage{amsfonts,amsmath,amssymb}
\usepackage{graphicx}
\usepackage{wrapfig}
\usepackage{subcaption}
\usepackage[colorlinks=true,bookmarks=false]{hyperref}
\usepackage{xcolor}
\definecolor{darkblue}{rgb}{0,0,0.45}
\definecolor{darkred}{rgb}{0.6,0,0}
\definecolor{darkgreen}{rgb}{0.13,0.5,0}
\hypersetup{colorlinks, linkcolor=darkblue, citecolor=darkgreen,urlcolor=darkblue}
\usepackage{algorithm}
\usepackage{algpseudocode} 
\usepackage{caption}

\let\oldep\everypar  \newtoks\everypar  \oldep{\the\everypar\looseness=-1}

\let\N=\bbbn
\let\R=\bbbr
\newcommand{\abs}[1]{\lvert#1\rvert}
\newcommand{\dist}{\mathrm{dist}}
\newcommand{\integers}[1]{\langle#1\rangle}

\newcommand{\classW}[1]{\mathsf{W[#1]}}
\newcommand{\classFPT}{\mathsf{FPT}}
\newcommand{\classNP}{\mathsf{NP}}
\newcommand{\classP}{\mathsf{P}}
\newcommand{\OPT}{\mathrm{OPT}}
\DeclareMathOperator{\cost}{cost}
\newcommand{\bigO}{\mathcal{O}}
\newcommand{\problemName}[1]{\textsc{#1}}

\newcommand{\tw}{\mathrm{tw}}
\let\hd=h
\let\dd=\Delta

\newcommand{\ball}[2]{
  B(#1, #2)
}

\newcommand{\problemkCenter}{\problemName{\(k\)\nobreakdash-Center}}
\newcommand{\problemkMedian}{\problemName{\(k\)\nobreakdash-Median}}

\newcommand{\problemkSupplier}{\problemName{\(k\)\nobreakdash-Supplier}}
\newcommand{\problemkCwO}{\problemName{\(k\)\nobreakdash-Center with Outliers}}
\newcommand{\problemkSwO}{\problemName{\(k\)\nobreakdash-Supplier with Outliers}}
\newcommand{\problemCkC}{\problemName{Capacitated \(k\)\nobreakdash-Center}}
\newcommand{\problemCkS}{\problemName{Capacitated \(k\)\nobreakdash-Supplier}}
\newcommand{\problemCkSwO}{\problemName{Capacitated \(k\)\nobreakdash-Supplier with Outliers}}

\newcommand{\Vclients}{V_C}
\newcommand{\Vsuppliers}{V_S}
\let\vcust=\Vclients
\let\vsup=\Vsuppliers
\let\outlier=\bot

\newcommand{\dl}{\mathrm{dl}}

\begin{document}

\title{%
  Generalized \(k\)-Center:\\
  \mbox{
    Distinguishing Doubling and Highway Dimension\thanks{%
      An extended abstract of this paper appeared at the 48th International Workshop on Graph-Theoretic Concepts in Computer Science in 2022.
    }
    \thanks{%
      This is the author’s version of a paper published in Algorithmica. The final version is available at \href{https://doi.org/10.1007/s00453-025-01357-1}{\texttt{https://doi.org/10.1007/s00453-025-01357-1}}.
    }
  }
}
\author{%
  Andreas Emil Feldmann \orcidID{0000-0001-6229-5332} \and~\newline Tung Anh Vu\(^\star\)\orcidID{0000-0002-8902-5196}
}
\institute{%
  Faculty of Mathematics and Physics, Charles University, Czech Republic\\
  \email{feldmann.a.e@gmail.com, tung@iuuk.mff.cuni.cz\(^\star\)}
}
\maketitle

\begin{abstract}
  We consider generalizations of the \problemName{\(k\)\nobreakdash-Center} problem in graphs of low doubling and highway dimension.
  For the \problemName{Capacitated \(k\)\nobreakdash-Supplier with Outliers (CkSwO)} problem, we show an efficient parameterized approximation scheme (EPAS) when the parameters are~\(k\), the number of outliers and the doubling dimension of the graph induced by the supplier set.
  On the other hand, we show that for the \problemName{Capacitated 
\(k\)\nobreakdash-Center} problem, which is a special case of 
\problemName{CkSwO}, obtaining a parameterized approximation scheme (PAS) 
is~\(\classW{1}\)\nobreakdash-hard when the parameters are~\(k\), and the 
highway dimension.
  This is the first known example of a problem for which it is hard to obtain a PAS for highway dimension, while simultaneously admitting an EPAS for doubling dimension.

\end{abstract}

\section{Introduction}

The well-known \problemkCenter{} problem and its generalizations has plenty of 
applications, for example 
selecting suitable locations for building hospitals to serve households of a 
municipality (see~\cite{AHMADIJAVID2017223} for a survey of healthcare 
facility location in practice).
In this setting, the number of hospitals we can actually build is limited, 
e.g.~by budgetary constraints.
We want to choose the locations so that the quality of the provided 
service is optimal, and 
a societally responsible way of measuring the quality of service is to ensure 
some minimal availability of healthcare to every household.
We can quantify this by measuring the distance of a household to its nearest 
hospital, and then minimize this distance over all households.
This strategy, however, does not account for the reality that healthcare 
providers have (possibly different) limits on the number of patients they can 
serve, and thus we introduce capacity constraints.
Furthermore, as the instances are given by transportation networks,
we model them by the titular doubling dimension and highway dimension, which we define later.


We formalize the problem as follows.
In the \problemCkS{} \problemName{(CkS)} problem, the input consists of a graph~\({G = (V, E)}\) with positive edge lengths, a set~\({\Vsuppliers \subseteq V}\) of \emph{suppliers}, a set~\({\Vclients \subseteq V}\) of \emph{clients}, a \emph{capacity function~\({L \colon \Vsuppliers \to \N}\)}, and an integer~\({k \in \N}\).
A \emph{feasible solution} is an \emph{assignment function~\({\phi \colon \Vclients \to \Vsuppliers}\)} such that~\({\abs{\phi(\Vclients)} \leq k}\) and for every supplier~\({u \in \phi(\Vclients)}\) we have~\({\abs{\phi^{-1}(u)} \leq L(u)}\).
For a pair of vertices~\({u, v \in V}\) we denote by~\({\dist_G(u, v)}\) the \emph{shortest-path distance} between vertices~\(u\) and~\(v\) with respect to edge lengths of~\(G\).
For a subset of vertices~\({W \subseteq V}\) and a vertex~\({u \in V}\), we 
denote~\({\dist_G(u, W) = \min_{w \in W} \dist_G(u, w)}\).
We omit the subscript $G$ if the graph is clear from context.
The \emph{cost} of a solution~\(\phi\) is defined as~\(\cost(\phi) = \max_{u \in \Vclients} \dist(u, \phi(u))\) and we want to find a feasible solution of minimum cost.
%
%
Let us mention the following special cases of \problemName{CkS}.
If~\({V = \Vsuppliers = \Vclients}\), then the problem is called \problemCkC{} \problemName{(CkC)}.
Additionally for \problemName{CkS} and \problemName{CkC} if~\({L(u) = \infty}\) for every supplier~\({u \in \Vsuppliers}\), then the problems are called \problemkSupplier{} and \problemkCenter{} respectively.
It is known that \problemkCenter{} is already \(\classNP\)\nobreakdash-hard~\cite{hochbaum1985best}.

Two popular approaches of dealing with \(\classNP\)\nobreakdash-hard problems are \emph{approximation algorithms}~\cite{vazirani2013approximation,williamson2011design} and \emph{parameterized algorithms}~\cite{10.1007/978-3-319-21275-3}.
Given an instance~\(\mathcal{I}\) of some minimization problem, 
a~\emph{\(c\)\nobreakdash-approximation algorithm} computes in polynomial time a 
solution of cost at most~\(c \cdot \OPT(\mathcal{I})\) 
where~\(\OPT(\mathcal{I})\) is the optimum cost of the instance~\(\mathcal{I}\), 
and we say that~\(c\) is the \emph{approximation ratio} of the algorithm.
If the instance is clear from context, we write only~\(\OPT\).
In a \emph{parameterized problem}, the input~\(\mathcal{I}\) comes with a \emph{parameter~\(q \in \N\)}.
If there exists an algorithm which computes the optimum solution in time~\(f(q) 
\cdot \abs{\mathcal{I}}^{\bigO(1)}\) where~\(f\) is some computable function, 
then we call such a problem \emph{fixed parameter tractable~(FPT)} and the 
algorithm an \emph{FPT algorithm}.
The rationale behind parameterized algorithms is to capture the ``difficulty'' of the instance by the parameter~\(q\) and then design an algorithm which is allowed to run in time superpolynomial in~\(q\) but retains a polynomial running time in the size of the input.
In this work, we focus on the superpolynomial part of the running time of 
\(\classFPT\) algorithms, so we will express~\(f(q) \cdot 
\abs{\mathcal{I}}^{\bigO(1)}\) as~\(\bigO^*(f(q))\);
in particular the ``\(\bigO^*\)'' notation ignores the polynomial factor in the input size.

It is known that unless~\(\classP = \classNP\), \problemkCenter{} and \problemkSupplier{} do not admit approximation algorithms with an approximation ratio better than~2 and~3~\cite{10.1145/5925.5933}.
It is shown in the same work that these results are tight by giving 
corresponding approximation algorithms.
For \problemCkC{}, An et al.~\cite{An2015-sr} give a~9\nobreakdash-approximation algorithm, and Cygan et al.~\cite{cygan2012lp} show a lower bound\footnote{%
  To the best of our knowledge, this reduction does not give a straightforward hardness result in the parameterized setting.%
} of~$3-\varepsilon$ for approximation assuming~\(\classP \neq \classNP\).
From the perspective of parameterized algorithms, Feldmann and 
Marx~\cite{feldmann2020parameterized} show that \problemkCenter{} 
is~\(\classW{1}\)\nobreakdash-hard in planar graphs of constant doubling 
dimension when the parameters are~\(k\), highway dimension and pathwidth.
Under the standard assumption~\(\classFPT \subsetneq \classW{1} \subsetneq 
\classW{2}\), this means that an \(\classFPT\) algorithm for \problemkCenter{} 
in planar constant doubling dimension graphs with the aforementioned parameters 
is unlikely to exist.
To overcome these hardness results, we will design \emph{parameterized 
\(c\)\nobreakdash-approximation algorithms}, which are algorithms with FPT 
runtime which output a solution of cost at most~\(c \cdot \OPT\).
The approach of parameterized approximation algorithms has been studied before, 
see the survey in~\cite{feldmann2020survey}.

Let us discuss possible choices for a parameter for these problems.
An immediate choice would be the size~\(k\) of the desired solution.
Unfortunately, Feldmann~\cite{feldmann2019fixed} has shown that approximating 
\problemkCenter{} within a ratio better than~2 when the parameter is~\(k\) 
is~\(\classW{2}\)\nobreakdash-hard.
So to design parameterized approximation algorithms, we must explore other parameters.
Guided by the introductory example, we focus on parameters which capture properties of transportation networks.

Abraham et al.~\cite{abraham2010highway} introduced the \emph{highway dimension} in order to explain fast running times of various shortest-path heuristics in road networks.
The definition of highway dimension is motivated by the following empirical observation of Bast et al.~\cite{bast2006transit,bast2007transit}.
Imagine we want to travel from some point~\(A\) to some sufficiently far point~\(B\) along the quickest route.
Then the observation is that if we travel along the quickest route, we will inevitably pass through a sparse set of ``access points''.
Highway dimension measures the sparsity of this set of access points around any vertex of a graph.
We give one of the several formal definitions of highway dimension, see~\cite{blum:LIPIcs:2019:11465,feldmann20181+varepsilon}.
Let~\((X, \dist)\) be a metric, for a point~\(u \in X\) and a radius~\(r \in \R^+\) we call the set~\(\ball{u}{r} = \{v \in X \mid \dist(u, v) \leq r\}\) the \emph{ball of radius~\(r\) centered at~\(u\)}.
\begin{definition}[{\cite{feldmann20181+varepsilon}}]
  \label{defi:hd1}
  The \emph{highway dimension} of a graph~\(G\) is the smallest integer~\(\hd\) such that, for some universal constant\footnote{See~\cite[Section 9]{feldmann20181+varepsilon} for a discussion. In essence, the highway dimension of a given graph can vary depending on the selection of~\(\gamma\).}~\(\gamma \geq 4\), for every~\(r \in \R^+\), and every ball~\(\ball{v}{\gamma r}\) of radius~\(\gamma r\) where~\(v \in V(G)\), there are at most~\(h\)~vertices in~\(\ball{v}{\gamma r}\) hitting all shortest paths of length more than~\(r\) that lie in~\(\ball{v}{\gamma r}\).
\end{definition}

%
We show the following hardness of parameterized approximation for 
\problemName{CkC} in low highway dimension graphs.
Among the definitions of highway dimensions, the one we use gives us the 
strongest hardness result, 
cf.~\cite{blum:LIPIcs:2019:11465,feldmann20181+varepsilon}.
\begin{theorem}
  \label{thm:hd1-hardness-of-constant-approximation}
  Consider any universal constant~\(\gamma\) in Definition~\ref{defi:hd1}.
  For any~\(\varepsilon > 0\), there is no parameterized \(((1 + \frac{1}{\gamma}) - \varepsilon)\)\nobreakdash-approximation algorithm for \problemName{CkC} with parameters~\(k\), treewidth\footnote{See Section~\ref{sec:kswo-epas-tw} for a formal definition.}, and highway dimension unless~\(\classFPT = \classW{1}\).
\end{theorem}

Another parameter we consider is \emph{doubling dimension}, defined as follows.
\begin{definition}
  The \emph{doubling constant} of a metric space~\((X, \dist)\) is the smallest value~\(\lambda\) such that for every~\(x \in X\) and every radius~\(r \in \R^+\), there exist at most~\(\lambda\)~points~\(y_1, \ldots, y_\lambda \in X\) such that~\(\ball{u}{r} \subseteq \cup_{i = 1}^\lambda \ball{y_i}{\frac{r}{2}}\).
  We say that the ball~\(\ball{x}{r}\) is \emph{covered} by 
balls~\(\ball{y_1}{\frac{r}{2}}, \ldots, \ball{y_\lambda}{\frac{r}{2}}\).
  The \emph{doubling dimension} $\dd(X)$ of~\(X\) is defined 
as~\(\log_2(\lambda)\).
The doubling dimension of a graph is the doubling dimension of its shortest 
path metric.
\end{definition}

Folklore results show that every metric for which the distance function is given by the~\(\ell_q\)\nobreakdash-norm in~\(D\)\nobreakdash-dimensional space~\(\R^D\) has doubling dimension~\(\bigO(D)\).
As a transportation network is embedded on a large sphere (namely the Earth), a reasonable model is to assume that the shortest-path metric abides to the Euclidean~\(\ell_2\)\nobreakdash-norm.
Buildings in cities form city blocks, which form a grid of streets.
Therefore it is reasonable to assume that the distances in cities are given by the Manhattan~\(\ell_1\)\nobreakdash-norm.
Road maps can be thought of as a mapping of a transportation network into~\(\R^2\).
It is then sensible to assume that transportation networks have constant doubling dimension.

Prior results on problems in graphs of low doubling and highway dimension went 
``hand in hand'' in the following sense.
For the \problemkMedian{} problem parameterized by the doubling dimension, 
Cohen-Addad et al.~\cite{cohenaddad2021kmediandd} show an \emph{efficient 
parameterized approximation scheme (EPAS)}, which is a parameterized 
algorithm that for some parameter~\(q\) and any~\(\varepsilon > 0\) 
outputs a solution of cost at most~\((1 + \varepsilon)\OPT\) and runs in 
time~\(\bigO^*(f(q, \varepsilon))\) where~\(f\) is a computable function.
In graphs of constant highway dimension, Feldmann and 
Saulpic~\cite{FELDMANN202172} follow up with a \emph{polynomial time 
approximation scheme (PTAS)} for \problemkMedian{}.
If we allow~\(k\) as a parameter as well, then Feldmann and 
Marx~\cite{feldmann2020parameterized} show an EPAS for \problemkCenter{} in low 
doubling dimension graphs, while Becker et 
al.~\cite{becker_et_al:LIPIcs:2018:9471} show an EPAS for \problemkCenter{}
in low highway dimension graphs.
By using the result of Talwar~\cite{talwar2004embedding} one can obtain 
\emph{quasi\nobreakdash-polynomial time approximation schemes \mbox{(QPTAS)}} 
for problems such as TSP, \problemName{Steiner Tree}, and \problemName{Facility 
Location} in low doubling dimension graphs.
Feldmann et al.~\cite{feldmann20181+varepsilon} extend this result to low highway dimension graphs and obtain analogous QPTASs.
Very recently, Feldmann and Filtser~\cite{tsp-ptas-hd} achieved a EPAS for \problemName{(Subset) TSP} in low highway dimension graphs by building on the PTAS for TSP in low doubling dimension graphs of Bartal, Gottlieb, and Krauthgamer~\cite{tsp-ptas-dd}.
The takeaway of this paragraph, based on the recurring trend, is that if one wants to design a PTAS for a certain problem in low highway dimension graphs, then one should start from a PTAS in the low doubling dimension setting and then attempt to extend it to the low highway dimension setting.
And by contrapositive with Theorem~\ref{thm:hd1-hardness-of-constant-approximation} in mind, we should expect that \problemName{CkC} is also hard in graphs of low doubling dimension.



Our main contribution lies in breaking the status quo by showing an EPAS for \problemName{CkC} in low doubling dimension graphs.
This is the first example of a problem, for which we provably cannot extend an 
algorithmic result in low doubling dimension graphs to the setting of low 
highway dimension graphs.\footnote{We remark that for this distinction to work, 
one has to be careful of the used definition of highway dimension: a stricter 
definition of highway dimension from~\cite{abraham2016highway} already implies 
bounded doubling dimension. On the other hand, for certain types of 
transportation networks, it can be argued that the doubling dimension is large, 
while the highway dimension is small. See~\cite[Appendix~A]{FELDMANN202172} 
for a detailed discussion.}
In fact, our algorithm even works in the supplier with outliers 
regime, where we are allowed to ignore some clients:
in the \problemCkSwO{} \problemName{(CkSwO)} problem, in addition to the \problemName{CkS} input~\((G, k, L)\), we are given an integer~\(p\).
A \emph{feasible solution} is an \emph{assignment}~\(\phi \colon \Vclients \to \Vsuppliers \cup \{\outlier\}\) which, in addition to the conditions specified in the definition of \problemName{CkS}, satisfies~\(\abs{\phi^{-1}(\outlier)} \leq p\).
Vertices~\(\phi^{-1}(\outlier)\) are called \emph{outliers}.
The goal is to find a solution of minimum cost, which is defined as~\(\cost(\phi) = \max_{u \in \Vclients \setminus \phi^{-1}(\outlier)} \dist(u, \phi(u))\).
Facility location and clustering with outliers were introduced by Charikar et al.~\cite{charikar2001algorithms}.
Among other results, they showed a 3\nobreakdash-approximation algorithm for 
\problemkCwO{} and an approximation lower bound of~$2-\varepsilon$.
Later, Harris et al.~\cite{harris2019lottery} and Chakrabarty et al.~\cite{chakrabarty2020nonuniform} independently closed this gap and showed a 2\nobreakdash-approximation algorithm for the problem.
For \problemName{CkSwO}, Cygan and Kociumaka~\cite{cygan_et_al:LIPIcs:2014:4462} show a 25\nobreakdash-approximation algorithm.
It may be of interest that the algorithm we show requires only that the doubling dimension of the graph induced by the supplier set to be bounded.
\begin{theorem}
  \label{thm:ckswo-epas-dd}
  Let~\(\mathcal{I} = (G, k, p, L)\) be an instance of \problemName{Capacitated \(k\)\nobreakdash-Supplier with Outliers}.
  Moreover, let~\((\Vsuppliers, \dist)\) be the shortest\nobreakdash-path metric induced by~\(\Vsuppliers\) and~\(\dd\) be its doubling dimension.
  There exists an algorithm which for any~\(\varepsilon > 0\) outputs a solution of cost~\((1 + \varepsilon)\OPT(\mathcal{I})\) in time~\(\bigO^*((k + p)^k \cdot \varepsilon^{-\bigO(k\dd)})\).
\end{theorem}

In light of the following results, this algorithm is almost the best we can hope for.
We have already justified the necessity of approximation by the result of Feldmann and Marx~\cite{feldmann2020parameterized}.
An EPAS parameterized only by~\(\dd\) is unlikely to exist, as Feder and Greene~\cite{10.1145/62212.62255} have shown that unless~\({\classP = \classNP}\), approximation algorithms with ratios better than~\(1.822\) and~2 for two\nobreakdash-dimensional Euclidean, resp.~Manhattan metrics cannot exist.
Hence it is necessary to parameterize by both~\(k\) and~\(\dd\).
The only improvement we can hope for is a better dependence on the number of outliers in the running time, e.g.~by giving an algorithm which is polynomial in~\(p\).

Given our hardness of approximation result for \problemName{CkC} on low highway 
dimension graphs in Theorem~\ref{thm:hd1-hardness-of-constant-approximation} and 
the known EPAS for \problemkCenter{} given by Becker et 
al.~\cite{becker_et_al:LIPIcs:2018:9471}, it is evident that the hardness stems 
from the introduction of capacities. For low doubling dimension graphs we were 
able to push the existence of an EPAS further than just introducing capacities, 
by considering suppliers and outliers. It therefore becomes an interesting 
question whether we can show an EPAS also for low highway dimension graphs when 
using suppliers and outliers, but without using capacities. The following 
theorem shows that this is indeed possible.

\begin{theorem}
  \label{thm:kswo-epas-hd}
  Let~\(\mathcal{I} = (G, k, p)\) be an instance of the \problemkSwO{} problem.
  There exists an EPAS for this problem with parameters~\(k\), \(p\), \(\varepsilon\), and highway dimension of~\(G\).
\end{theorem}

\subsection{Used techniques}
\label{subsec:used-techniques}

To prove Theorem~\ref{thm:hd1-hardness-of-constant-approximation}, we enhance a result of Dom et al.~\cite{10.1007/978-3-540-79723-4_9} which shows that \problemName{Capacitated Dominating Set} is~\(\classW{1}\)\nobreakdash-hard in low treewidth graphs by reducing from a~\(\classW{1}\)\nobreakdash-hard problem.
In fact, the graph we produce in our reduction has the exact same structure as~\cite{10.1007/978-3-540-79723-4_9}.
We prove that if we set the graph's edge weights in a certain way, then its highway dimension will be polynomial in~\(k\).

We prove Theorem~\ref{thm:ckswo-epas-dd} by using the concept of a~\(\delta\)\nobreakdash-net which is a sparse subset of the input metric such that every input point has a net point near it.
In Lemma~\ref{lem:ckswo-delta-net-upper-bound} we show that the size of the net, for~\(\delta = \varepsilon \cdot \OPT\), can be bounded by parameters~\(k\), number of outliers~\(p\), doubling dimension, and~\(\varepsilon\).
A naive approach, which ignores capacities of suppliers, would be to guess a \(k\)\nobreakdash-size subset of the net which is a feasible~\((1 + \varepsilon)\)\nobreakdash-approximate solution.
This is in fact the approach used by Feldmann and Marx~\cite{feldmann2020parameterized} to show an EPAS for \problemkCenter{} in low doubling dimension graphs.
To support the capacities, we need further ideas regarding the structure of the solution with respect to the net.
We present these in Lemmas~\ref{lem:recognize-feasible-ckswo-solution} and~\ref{lem:ckswo-epas-dd-main}.

To prove Theorem~\ref{thm:kswo-epas-hd}, we generalize the EPAS for \problemkCenter{} in low highway dimension graphs by Becker et al.~\cite{becker_et_al:LIPIcs:2018:9471}.
A major component of this algorithm is an EPAS for \problemName{kSwO} in low treewidth graphs, cf.~Theorem~\ref{thm:kswo-epas-tw}, which generalizes an EPAS for \problemkCenter{} in low treewidth graphs by Katsikarelis et al.~\cite{KATSIKARELIS201990}.

\section{Inapproximability in low highway dimension graphs}
\label{sec:hd-hardness}

In this section we are going to prove Theorem~\ref{thm:hd1-hardness-of-constant-approximation}, i.e.~we show that there is no parameterized approximation scheme for \problemName{Capacitated \(k\)\nobreakdash-Center} in graphs of low highway dimension unless~\(\classFPT = \classW{1}\).

We reduce from \problemName{Multicolored Clique}, which is known to be a~\(\classW{1}\)\nobreakdash-hard problem~\cite[Theorem~13.25]{10.1007/978-3-319-21275-3}.
The input of \problemName{Multicolored Clique} consists of a graph~\(G\) and an integer~\(k\).
The vertex set of~\(G\) is partitioned into \emph{color classes~\(V_1, \ldots, V_k\)} where each color class is an independent set.
The goal is to find a \(k\)\nobreakdash-clique.
Note that endpoints of every edge of~\(G\) have to be in different color classes.
Hence if a \(k\)\nobreakdash-clique exists in~\(G\), then it has exactly one vertex in each color class.

To prove Theorem~\ref{thm:hd1-hardness-of-constant-approximation}, we will need several settings of edge lengths.
Namely for every~\(\lambda \geq 2\gamma \geq 8\), given an instance~\(\mathcal{I} = (G, k)\) of \problemName{Multicolored Clique}, we produce in polynomial time a \problemName{CkC} instance~\(\overline{\mathcal{I}}_\lambda = (\overline{G}, \overline{k}, L, d_\lambda)\) where the highway dimension and treewidth of~\(\overline{G}\) is~\(\bigO(k^4)\) and~\({\overline{k} = 7k(k - 1) + 2k}\).
If we are not interested in a particular setting of~\(\lambda\) or we speak generally about all instances for all possible settings of~\(\lambda\), we omit the subscript.

\begin{figure}[b]
  \centering
  \includegraphics{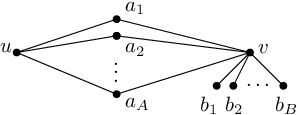}
  \caption{The result of adding an~\((A, B)\)\protect\nobreakdash-arrow from~\(u\) to~\(v\).}
  \label{fig:ab-arrow}
\end{figure}

It follows from~\cite{10.1007/978-3-319-21275-3} that we can assume without loss of generality that every color class consists of~\(N\)~vertices and the number of edges between every two color classes is~\(M\).
For two integers~\(m \leq n\) by~\(\integers{m, n}\) we mean the set of integers~\(\{m, m+1, \ldots, n\}\), and~\(\integers{m} = \integers{1, m}\).
For distinct~\(i, j \in \integers{k}\) we denote by~\(E_{i,j}\) the set of ordered pairs of vertices~\((u, v)\) such that~\(u \in V_i\), \(v \in V_j\), and~\(\{u, v\}\) is an edge in~\(G\).
When we add an~\emph{\((A, B)\)\nobreakdash-arrow} from vertex~\(u\) to vertex~\(v\), we add~\(A\)~subdivided edges between~\(u\) and~\(v\) and additionally we add~\(B\) unique vertices to the graph and connect them to~\(v\), see Fig.~\ref{fig:ab-arrow}.
When we \emph{mark} a vertex~\(u\), we add~\(\overline{k} + 1\) new vertices to the graph and connect them to~\(u\).
We denote the set of all marked vertices by~\(Z\).

\begin{figure}[t]
  \centering
  \includegraphics{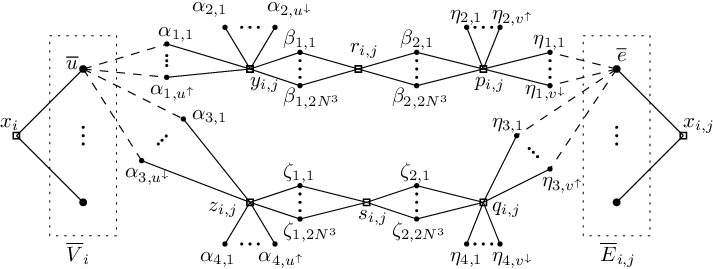}
  \caption{%
    Part of the reduction for color class~\(V_i\) and edge set~\(E_{i, j}\).
    Vertex~\(\overline{e}\) represents an edge~\({(v, w) \in E_{i, j}}\) in~\(G\).
    We omit the sets~\(S_i\) and~\(S_{i, j}\).
    Marked vertices are drawn by boxes and we omit their~\(\overline{k} + 1\)~``private'' neighbors.
    Edges drawn by a dashed line have length~\(1\) and the remaining edges have length~\(\lambda\).
    We also omit the appropriate arrows from vertices of~\(\overline{V}_i \setminus \{\overline{u}\}\) to~\(y_{i, j}\) and to~\(z_{i, j}\), and the appropriate arrows from vertices of~\(\overline{E}_{i, j} \setminus \{\overline{e}\}\) to~\(p_{i, j}\) and to~\(q_{i, j}\).
  }
  \label{fig:reduction}
\end{figure}

We first describe the structure of~\(\overline{G}\) and we set the capacities and edge lengths of~\(\overline{G}\) afterwards.
See Fig.~\ref{fig:reduction} for an illustration of the reduction.
As we have mentioned in Section~\ref{subsec:used-techniques}, the structure of~\(\overline{G}\) is exactly the same as in~\cite{10.1007/978-3-540-79723-4_9}.

\paragraph{Color class gadget.}
For each color class~\(V_i\), we create a gadget as follows.
We arbitrarily order vertices of~\(V_i\) and to the~\(j\textsuperscript{th}\) vertex~\(u \in V_i\) we assign numbers~\(u^\uparrow = j \cdot 2N^2\) and~\(u^\downarrow = 2N^3 - u^\uparrow\).
For each vertex~\(u \in V_i\) we create a vertex~\(\overline{u}\) and we denote~\(\overline{V}_i = \{\overline{u} \mid u \in V_i\}\).
We add a marked vertex~\(x_i\) and connect it to every vertex of~\(\overline{V}_i\).
We add a set~\(S_i\) of~\(\overline{k} + 1\)~vertices and connect each vertex of~\(S_i\) to every vertex of~\(\overline{V}_i\).
For every~\(j \in \integers{k} \setminus \{i\}\) we add a pair of marked vertices~\(y_{i, j}\) and~\(z_{i, j}\).
We denote~\(Y_i = \cup_{j \in \integers{k} \setminus \{i\}} \{y_{i, j}, z_{i, j}\}\).
For every vertex~\(\overline{u} \in \overline{V}_i\) we add a~\((u^{\uparrow}, u^{\downarrow})\)-arrow from~\(u\) to each vertex of~\(\cup_{j \in \integers{k} \setminus \{i\}} y_{i, j}\) and a~\((u^{\downarrow}, u^{\uparrow})\)-arrow from~\(u\) to each vertex of~\(\cup_{j \in \integers{k} \setminus \{i\}} z_{i, j}\).

\paragraph{Edge set gadget.}
For every~\(i \in \integers{k - 1}, j \in \integers{i + 1, k}\) we create a gadget for the edge set~\(E_{i, j}\) as follows.
For every edge~\(e \in E_{i, j}\) we create a vertex~\(\overline{e}\) and we denote~\(\overline{E}_{i, j} = \{\overline{e} \mid e \in E_{i, j}\}\).
We add a marked vertex~\(x_{i, j}\) and connect it to every vertex of~\(\overline{E}_{i, j}\).
We add a set~\(S_{i, j}\) of~\(\overline{k} + 1\)~vertices and connect each vertex of~\(S_{i, j}\) to every vertex of~\(\overline{E}_{i, j}\).
We add four marked vertices~\(p_{i, j}, p_{j, i}, q_{i, j},q_{j, i}\).
Consider an edge~\(e = (u, v) \in E_{i, j}\), we connect~\(\overline{e}\) to~\(p_{i, j}\) with a~\((u^\downarrow, u^\uparrow)\)-arrow, to~\(q_{i, j}\) with a~\((u^\uparrow, u^\downarrow)\)-arrow, to~\(p_{j, i}\) with a~\((v^\downarrow, v^\uparrow)\)-arrow, and to~\(q_{j, i}\) with a~\((v^\uparrow, v^\downarrow)\)-arrow.
We denote~\(\mathcal{S} = (\cup_{i \in \integers{k}} S_i) \cup (\cup_{i \in \integers{k - 1}, j \in \integers{i + 1, k}} S_{i, j})\).

\paragraph{Adjacency gadget.}
To connect the color class gadgets and the edge set gadgets, for every distinct~\(i, j \in \integers{k}\) we add marked vertices~\(r_{i, j}\) and~\(s_{i, j}\), and we add~\((2N^3, 0)\)-arrows from~\(y_{i, j}\) to~\(r_{i, j}\), from~\(p_{i, j}\) to~\(r_{i, j}\), from~\(z_{i, j}\) to~\(s_{i, j}\), and from~\(q_{i, j}\) to~\(s_{i, j}\).

\paragraph{Capacities.}
We now describe the capacities~\({L \colon V(\overline{G}) \to \N}\).
To streamline the exposition, we will assume that each vertex of~\(\phi(V(\overline{G}))\) covers itself at no ``cost'' with respect to the capacity.
For every two distinct~\({i, j \in \integers{k}}\), the vertex~\(x_i\) has capacity~\({N - 1 + \overline{k} + 1}\), the vertex~\(x_{i, j}\) has capacity~\({M - 1 + \overline{k} + 1}\), vertices~\(y_{i, j}\) and~\(z_{i, j}\) have capacity~\({2N^4 + \overline{k} + 1}\), vertices~\(p_{i, j}\) and~\(q_{i, j}\) have capacity~\({2MN^3 + \overline{k} + 1}\), vertices~\(r_{i, j}\) and~\(s_{i, j}\) have capacity~\({2N^3 + \overline{k} + 1}\), and the remaining vertices have capacity equal to their degree.

\paragraph{Edge lengths.}
Given~\(\lambda \in \R^+\) we set edge lengths~\(d_\lambda \colon E(\overline{G}) \to \R\) as follows.
For every~\(i \in \integers{k}\) and every vertex~\(\overline{u} \in \overline{V}_i\), we assign length~\(1\) to edges between~\(\overline{u}\) and~\(N(\overline{u}) \setminus (\{x_i\} \cup S_i)\), i.e.~the set of vertices originating from subdivided edges of arrows between~\(\overline{u}\) and~\(Y_i\).
Similarly for every~\(i \in \integers{k - 1}, j \in \integers{i + 1, k}\) and every vertex~\(\overline{v} \in \overline{E}_{i, j}\), we assign length~\(1\) to edges between~\(\overline{v}\) and~\(N(\overline{v}) \setminus (\{x_{i, j}\} \cup S_{i, j})\).
To the remaining edges we assign length~\(\lambda\).

From the way we assign edge lengths~\(d_\lambda\), in a solution~\(\phi\) of cost~\(\lambda + 1\) such that~\(\abs{\phi(V(\overline{G}))} \leq \overline{k}\), it must be the case that~\(Z \subseteq \phi(V(\overline{G}))\), since~\(Z\) is the set of marked vertices with~\(\overline{k} + 1\) private neighbors.

Dom et al.~\cite[Observation 1]{10.1007/978-3-540-79723-4_9} observe that~\(\tw(\overline{G}) = \bigO(k^4)\):
The size of~\(\mathcal{S}\) is~\(\bigO(k^4)\) and the size of~\(Z\) is~\(\frac{1}{2}(13k^2 - 11k)\).
Removing~\(\mathcal{S} \cup Z\) leaves us with a forest and the observation follows.

We prove that~\(\overline{G}\) has bounded highway dimension.
This is not immediate since treewidth and highway dimension are incomparable graph parameters, c.f.~\cite{blum:LIPIcs:2019:11465}.
For an interval of reals~\(I\), we will consider shortest paths of~\(\overline{G}\) such that their length belongs to~\(I\), thus we accordingly denote this set by~\(\mathcal{P}_I\).
We also denote~\(\mathcal{V} = (\cup_{i = 1}^k \overline{V_i}) \cup (\cup_{1 \leq i < j \leq k} \overline{E}_{i, j})\).
\begin{lemma}
  \label{lem:ckc-bounded-hd1}
  For any~\(\lambda \geq 2\gamma\), where~\(\gamma\) is the universal constant in Definition~\ref{defi:hd1}, graph~\(\overline{G}\) with edge lengths~\(d_\lambda\) has highway dimension~\(\hd(\overline{G}) \in \bigO(k^4)\).
\end{lemma}

\begin{proof}
  Consider a scale~\(r \geq 2\) and any shortest path~\(\pi \in \mathcal{P}_{(r, \infty)}\).
  Let~\({H = \mathcal{S} \cup Z}\).
  From the way we constructed~\(\overline{G}\), the maximal length of a shortest path which does not contain an edge of length~\(\lambda\) is~2.
  Thus~\(\pi\) has to contain an edge of length~\(\lambda\).
  Every edge of length~\(\lambda\) is adjacent to a vertex of~\(H\), which means that~\(H\) forms a hitting set for~\(\mathcal{P}_{(r, \infty)}\).
  We have~\(\abs{\mathcal{S} \cup Z} = \bigO(k^4)\), and this trivially bounds~\(\abs{\ball{u}{\gamma r} \cap H} = \bigO(k^4)\) for any vertex~\(u \in V(\overline{G})\).

  Now consider the case~\(r < 2\).
  Let~\(H = \mathcal{S} \cup \mathcal{V} \cup Z\).
  We have already seen that every edge of length~\(\lambda\) is incident to~\(\mathcal{S} \cup Z\).
  Every edge of length~\(1\) is incident to a vertex of~\(\mathcal{V}\), hence the set~\(H\) hits all shortest paths~\(\mathcal{P}_{(r, \infty)}\).
  We have~\(\gamma r < \lambda\), as~\(\lambda \geq 2\gamma\).
  This means that if~\(\ball{u}{\gamma r}\) contains a vertex~\(v\) other than~\(u\), then~\(\dist(u, v) < \lambda\).
  From the construction of the graph, it then must be the case that~\(\dist(u, v) = 1\), and in fact~\(\{u, v\} \in E(\overline{G})\) and~\(d_\lambda(u, v) = 1\).
  Let~\(F = \{e \in E(\overline{G}) \mid d_\lambda(e) = 1\}\).
  We have to bound the size of~\(\ball{u}{\gamma r} \cap H\) where~\(u\) is an endpoint of an edge in~\(F\).
  We will show that the bound is~\(1\).

  For every edge of~\(F\), one of its endpoints is a vertex~\(u \in \mathcal{V}\) and the other is a vertex which belongs to a vertex originating from a subdivided edge of an arrow incident to~\(u\).
  In the first case, a shortest path between any pair of vertices of~\(\mathcal{V}\) has to contain two edges of length~\(\lambda\).
  Thus the distance between them is at least~\(2\lambda\).
  Any shortest path from a vertex of~\(\mathcal{V}\) to a vertex of~\(\mathcal{S} \cup Z\) contains an edge of length~\(\lambda\).
  Thus for a vertex~\(u \in \mathcal{V}\), we have~\(\ball{u}{\gamma r} \cap H = \{u\}\) as~\(\gamma r < \lambda\).
  In the second case, a vertex~\(v\) belonging to a subdivided edge has degree two and exactly one of the edges incident to it has length~\(1\); the other has length~\(\lambda\), which is not consequential in the current case as~\(\gamma r < \lambda\).
  The neighbor~\(w\) of~\(v\) such that~\(d_\lambda(v, w) = 1\) belongs to~\(\mathcal{V}\).
  This means that~\({\abs{\ball{v}{\gamma r} \cap H} = 1}\) since~\({\ball{v}{\gamma r} = \{v, w\}}\), i.e.~we have considered all the possible cases.
  \qed
\end{proof}

Now we prove that~\(\mathcal{I}\) contains a~\(k\)\nobreakdash-clique if and only if~\(\overline{\mathcal{I}}_\lambda\) contains a solution of cost at most~\(\lambda + 1\).
As~\(d_\lambda\) assigns edge lengths~1 and~\(\lambda\), this will imply that if~\(\mathcal{I}\) does not contain a~\(k\)\nobreakdash-clique, then any solution of~\(\overline{\mathcal{I}}\) has to have cost at least~\(\lambda + 2\) and vice versa.

The forward implication follows implicitly from the original result of Dom et al.~\cite{10.1007/978-3-540-79723-4_9}, since we can interpret a capacitated dominating set as a solution of cost~\(\lambda\), hence we omit the proof.
\begin{lemma}[{\cite[Lemma~1]{10.1007/978-3-540-79723-4_9}}]
  \label{lem:mc-to-ckc}
  If~\(\mathcal{I}\) contains a~\(k\)\nobreakdash-clique, then~\(\overline{\mathcal{I}}_\lambda\) contains a solution of cost~\(\lambda\).
\end{lemma}

To prove the backward implication, we start need to show that a solution of cost~\(\lambda + 1\) has to open a vertex in every~\(\overline{V}_i\) and every~\(\overline{E}_{i, j}\).
In contrast to Lemma~\ref{lem:mc-to-ckc}, the backward implication does not simply follow from the original result since we have added edge lengths to the graph.
\begin{lemma}
  \label{lem:ckc-cost-lambda-plus-1-solution-appearance}
  Let~\(\phi\) be a solution of~\(\overline{\mathcal{I}}_\lambda\) of cost~\(\lambda + 1\), and~\(D = \phi(V(\overline{G}))\).
  Then for each~\(i \in \integers{k}\) we have~\(\abs{\overline{V}_i \cap D} = 1\) and for each~\(i \in \integers{k - 1}, j \in \integers{i + 1, k}\) we have~\(\abs{\overline{E}_{i, j} \cap D} = 1\).
\end{lemma}
\begin{proof}
  We prove this statement by contradiction.
  Suppose that there exists~\(i\) such that~\(\abs{\overline{V}_i \cap D} \neq 1\).
  We first consider the case when~\(\abs{\overline{V}_i \cap D} = 0\).
  Vertices of~\(S_i\) have to be covered by vertices at distance at most~\(\lambda + 1\) from them excluding those belonging to~\(\overline{V}_i\).
  Let~\(W\) be the set of vertices originating from subdivided edges of arrows between~\(\overline{V}_i\) and~\(Y_i\).
  Then it must be the case that~\(\phi^{-1}(S_i) \subseteq S_i \cup W\) as~\((\cup_{u \in S_i} \ball{u}{\lambda + 1}) \setminus \overline{V_i} = S_i \cup W\).
  The distance between a pair of vertices of~\(S_i\) is~\(2\lambda\), which is strictly greater than~\(\lambda + 1\) since~\(\lambda \geq 2\gamma \geq 8\).
  Hence for a vertex~\(u \in (D \cap S_i)\) we have~\(\abs{\phi^{-1}(u) \cap S_i} \leq 1\).
  For a vertex~\(w \in W\) we have~\(L(w) = 2\), and so~\(\abs{\phi^{-1}(w) \cap S_i} \leq 2\).
  In total, we would have to pick at least~\(\frac{\overline{k} + 1}{2}\) vertices of~\(S_i \cup W\) to cover~\(S_i\).
  We have~\(\abs{Z} > \overline{k} - \frac{\overline{k} + 1}{2}\) and~\(D\) must contain~\(Z\).
  As~\(Z \cap (S_i \cup W) = \emptyset\), this contradicts the fact that~\(\abs{D} \leq \overline{k}\) and thus~\(\abs{\overline{V}_i \cap D} \geq 1\).

  In the case that there exist~\(i\) and~\(j\) such that~\(\abs{\overline{E}_{i, j} \cap D} = 0\), we can apply a similar argument to show that~\(\abs{\overline{E}_{i, j} \cap D} \geq 1\)

  We have shown that every~\(\overline{V}_i\) and every~\(\overline{E}_{i, j}\) contains at least one vertex of~\(D\).
  By calculating~\(\overline{k} - \abs{Z} = k + \binom{k}{2}\), which is equal to the number of color classes plus the number of edge sets, it follows that each of them must contain exactly~\(1\)~vertex of~\(D\) as~\(Z \subset D\).
  \qed
\end{proof}

Now we show that if~\(\overline{\mathcal{I}}_\lambda\) contains a solution of cost~\(\lambda + 1\), then~\(\mathcal{I}\) contains a~\(k\)\nobreakdash-clique.
\begin{lemma}
  \label{lem:ckc-lambda-plus-1-to-clique}
  If~\(\overline{\mathcal{I}}_\lambda\) has a solution~\(\phi\) of cost~\(\lambda + 1\), then~\(\mathcal{I}\) contains a~\(k\)\nobreakdash-clique.
\end{lemma}
\begin{proof}
  Let~\(D = \phi(V(\overline{G}))\).
  For~\(i \in \integers{k}\) let~\(\overline{u}_i\) be the vertex of~\(\overline{V}_i \cap D\) and for~\({i \in \integers{k - 1}}, j \in \integers{i + 1, k}\) let~\(\overline{e}_{i, j}\) be the vertex of~\(\overline{E}_{i, j} \cap D\).
  These vertices are well-defined by Lemma~\ref{lem:ckc-cost-lambda-plus-1-solution-appearance}.
  To prove that these vertices encode a~\(k\)\nobreakdash-clique in~\(G\), we want to show for every~\(i \in \integers{k - 1}, j \in \integers{i + 1, k}\) that vertices~\(u_i\) and~\(u_j\), which correspond to~\(\overline{u}_i\) and~\(\overline{u}_j\) respectively, are incident to the edge~\(e_{i, j}\) corresponding to the vertex~\(\overline{e}_{i, j}\).
  We will only present the proof of incidence for~\(u_i\) and~\(e_{i, j}\), for~\(u_j\) we can proceed analogously.
  Let~\(v\) be the vertex of the edge~\(e_{i, j}\) which belongs to~\(V_i\) in~\(G\).
  Before we prove that~\(u_i\) and~\(e_{i, j}\) are incident, we first argue that~\(u_i^\uparrow + v^\downarrow = 2N^3\).

  We prove this statement by contradiction.
  First suppose~\(u_{i}^\uparrow + v^\downarrow < 2N^3\).
  Then~\(u_{i}^\uparrow + v^\downarrow \leq 2N^3 - 2N^2\) as for every two distinct vertices~\(w_1\) and~\(w_2\) of a color class we have~\(\abs{w_1^\uparrow - w_2^\uparrow} \geq 2N^2\) and~\(\abs{w_1^\downarrow - w_2^\downarrow} \geq 2N^2\).
  Consider the set
  \begin{equation}
    \mathcal{T} = (N(y_{i, j}) \cup N(r_{i, j}) \cup N(p_{i, j})) \setminus \phi^{-1}(\{x_i, u_i, e_{i, j}, x_{i, j}\}).
  \end{equation}
  It follows that in a solution of cost~\(\lambda + 1\), vertices of~\(\mathcal{T}\) must be covered by~\(y_{i, j}\), \(r_{i, j}\) or~\(p_{i, j}\) as edges of~\(\overline{G}\) have length~\(1\) or~\(\lambda\).
  We have~\(L(y_{i, j}) + L(r_{i, j}) + L(p_{i, j}) = 2N^4 + 2MN^3 + 2N^3 + 3(\overline{k} + 1)\).
  However,
  \begin{align}
    \abs{\mathcal{T}} & \geq 2N^4 + 2MN^3 + 4N^3 + 3(\overline{k} + 1) - ((2N^3 - 2N^2) + (N - 1) + (M - 1)) \notag\\
                      & > 2N^4 + 2MN^3 + 2N^3 + 3(\overline{k} + 1),
  \end{align}
  where we used that~\(M \leq N^2\) and~\(N^2 > N\).
  Thus~\(y_{i, j}\), \(r_{i, j}\), and~\(p_{i, j}\) cannot cover~\(\mathcal{T}\).
  This contradicts the fact that~\(\phi\) is a solution of cost~\(\lambda + 1\).

  If~\(u_i^\uparrow + v^\downarrow > 2N^3\), then~\(u_i^\downarrow + v^\uparrow < 2N^3\) and we can apply the identical argument for vertices~\(z_{i, j}, s_{i, j}, q_{i, j}\).

  It remains to prove that~\(u_i\) is incident to~\(e_{i, j}\).
  Again, let~\(v\) be the vertex of~\(e_{i, j}\) which lies in~\(V_i\) of~\(G\).
  We know from the preceding argument that~\(u_i^\uparrow + v^\downarrow = 2N^3\).
  However, the only vertex~\(w \in V_i\) such that~\(w^\downarrow = 2N^3 - u_i^\uparrow\) is~\(u_i\) itself;
  for any~\(w \in V_i \setminus \{u_i\}\) we would have~\(\abs{(u_i^\uparrow + w^\downarrow) - 2N^3} \geq 2N^2\).
  Hence~\(v = u_i\) and so~\(e_{i, j}\) is incident to~\(u_i\).
  This concludes the proof.
  \qed
\end{proof}

We are ready to prove Theorem~\ref{thm:hd1-hardness-of-constant-approximation}.
\begin{proof}[{of Theorem~\ref{thm:hd1-hardness-of-constant-approximation}}]
  Let~\(\gamma \geq 4\) be the universal constant in Definition~\ref{defi:hd1}, and~\(\lambda = 2\gamma\).
  For contradiction, suppose that there exists an algorithm~\(\mathcal{A}\) with an approximation ratio~\(c = \left((1 + \frac{1}{\gamma}) - \varepsilon\right)\) for some fixed~\(\varepsilon > 0\) parameterized by~\(k\), treewidth and highway dimension.

  Given a \problemName{Multicolored Clique} instance~\(\mathcal{I} = (G, k)\), we produce a \problemName{CkC} instance~\(\overline{\mathcal{I}}_\lambda = (\overline{G}, \overline{k}, L)\) with edge lengths~\(d_\lambda\) using the reduction from Section~\ref{sec:hd-hardness}.
  We run algorithm~\(\mathcal{A}\) on the instance~\(\mathcal{\overline{I}}\), which takes time~\(\bigO^*(f(\overline{k}, \tw(\overline{G}), \hd(\overline{G})))\) for some computable function~\(f\).
  We have~\(\overline{k} = 7k(k - 1) + 2k\).
  By the observation of Dom et al.~\cite{10.1007/978-3-540-79723-4_9}, graph~\(\overline{G}\) has treewidth~\(\bigO(k^4)\).
  By Lemma~\ref{lem:ckc-bounded-hd1}, graph~\(\overline{G}\) has highway dimension~\(\bigO(k^4)\).
  Thus there exists a computable function~\(g\) such that~\(f(\overline{k}, \tw(\overline{G}), \hd(\overline{G})) \leq g(k)\).
  If~\(\mathcal{I}\) is a YES\nobreakdash-instance, then~\(\mathcal{\overline{I}}_\lambda\) has a solution of cost~\(\lambda\) by Lemma~\ref{lem:mc-to-ckc}.
  On the other hand, if~\(\mathcal{I}\) is a NO\nobreakdash-instance, then any solution of~\(\mathcal{\overline{I}}\) has to have cost at least~\(\lambda + 2\), which follows from Lemma~\ref{lem:ckc-lambda-plus-1-to-clique}.
  Algorithm~\(\mathcal{A}\) is able to distinguish between these two cases in time~\(\bigO^*(f(\overline{k}, \tw(\overline{G}), \hd(\overline{G}))) = \bigO^*(g(k))\), since~\(c < \frac{\lambda + 2}{\lambda}\).
  This is an~\(\classFPT\)~algorithm for a~\(\classW{1}\)\nobreakdash-hard problem, which contradicts~\(\classFPT \neq \classW{1}\).
  \qed
\end{proof}

\section{EPAS on graphs of bounded doubling dimension}

In this section we prove Theorem~\ref{thm:ckswo-epas-dd}, i.e.~we show the existence of an EPAS for \problemName{CkSwO} on instances where the graph induced by the supplier set has bounded doubling dimension.
To be more precise, we develop a decision algorithm which, given a cost~\(\varrho \in \R^+\), and~\(\varepsilon > 0\), computes a solution of cost~\((1 + \varepsilon) \varrho\) in \(\classFPT\) time with parameters~\(k\), \(p\), doubling dimension and~\(\varepsilon\).
Formally, the result is the following lemma.

\begin{lemma}
  \label{lem:ckswo-epas-dd}
  Let~\(\mathcal{I} = (G, k, p, L)\) be a \problemName{CkSwO} instance.
  Moreover, let~\((\Vsuppliers, \dist)\) be the shortest\nobreakdash-path metric induced by~\(\Vsuppliers\) and~\(\dd\) be its doubling dimension.
  There exists an algorithm which, given a cost~\(\varrho \in \R^+\) and~\(\varepsilon > 0\), either
  \begin{itemize}
    \item computes a feasible solution of cost~\((1 + \varepsilon)\varrho\) if~\((1 + \varepsilon)\varrho \geq \OPT(\mathcal{I})\), or
    \item correctly decides that~\(\mathcal{I}\) has no solution of cost at most~\(\varrho\),
  \end{itemize}
  running in time~\(\bigO^*\left((k + p)^k\varepsilon^{-\bigO(k\dd)}\right)\).
\end{lemma}

Using Lemma~\ref{lem:ckswo-epas-dd}, we can obtain the algorithm of Theorem~\ref{thm:ckswo-epas-dd} as follows.
We can first assume without loss of generality, that~\(\Vclients \cup \Vsuppliers = V(G)\).
Suppose that we can guess the optimum cost~\(\OPT\) of any \problemName{CkSwO} instance.
By using~\(\OPT\) as~\(\varrho\) in Lemma~\ref{lem:ckswo-epas-dd}, we can output a solution of cost~\((1 + \varepsilon)\OPT\).
To guess the optimum cost~\(\OPT\), observe that~\(\OPT\) must be one of the inter-vertex distances.
Hence the minimum inter-vertex distance~\(\varrho\) for which the algorithm outputs a solution has the property that~\(\varrho \leq \OPT\) and consequently~\((1 + \varepsilon)\varrho \leq (1 + \varepsilon)\OPT\).

The main ingredient of the algorithm is the notion of a~\(\delta\)\nobreakdash-net.
For a metric~\((X, \dist)\), a subset~\(Y \subseteq X\) is called a \emph{\(\delta\)\nobreakdash-cover} if for every~\(u\in X\) there exists a~\(v \in Y\) such that~\(\dist(u, v) \leq \delta\).
If a \(\delta\)\nobreakdash-cover~\(Y\) has an additional property that for every two distinct~\(u, v \in Y\) we have~\(\dist(u, v) > \delta\), then we say that~\(Y\) is a \emph{\(\delta\)\nobreakdash-net}.
Observe that a~\(\delta\)\nobreakdash-net can be computed greedily in polynomial time.

Let us give the main idea behind the algorithm.
Given an instance of the problem and~\(\varepsilon > 0\), let~\(\phi^*\) be an optimum solution of cost~\(\OPT\), \(\Vclients^*\) be clients that are not outliers according to~\(\phi^*\), i.e.~\(\Vclients^* = \{u \in \Vclients \mid \phi^*(u) \neq \outlier\}\), and~\(Y\) be an~\((\varepsilon \cdot \OPT)\)\nobreakdash-net of the metric~\((\Vsuppliers, \dist)\).
Consider an assignment function~\(\phi\) constructed as follows.
For each client~\(u \in \Vclients^*\) we set~\(\phi(u)\) to the nearest point of~\(Y\) to~\(\phi^*(u)\), and for the remaining clients we set the value of~\(\phi\) to~\(\outlier\).
If for every selected supplier~\(s \in (\phi(\Vclients) \setminus \{\outlier\})\) we have~\(\abs{\phi^{-1}(s)} \leq L(s)\), then~\(\phi\) is a feasible solution.
Since~\(Y\) is an~\((\varepsilon \cdot \OPT)\)\nobreakdash-net, the cost of~\(\phi\) is at most~\((1 + \varepsilon)\OPT\).

The main obstacle to implementing an algorithm from this idea is that we do not know the optimum solution~\(\phi^*\).
However, by the definition of the net~\(Y\), we know that each selected supplier~\(\phi^*(\Vclients^*)\) is near some point of~\(Y\).
If~\(Y\) was not too large, we could guess which~\(k\)~of its points are near to every supplier of~\(\phi^*(\Vclients^*)\).
Later, we will also show how to ensure that the solution we create respects capacities of suppliers we pick.

We now show how to bound the size of the net.
Let~\((X, \dist)\) be a metric of doubling dimension~\(\dd\), by the \emph{aspect ratio} of a set~\(X' \subseteq X\), we mean the diameter of~\(X'\) divided by the minimum distance between any two distinct points of~\(X'\), that is~\(\frac{\max_{u, v \in X'} \dist(u, v)}{\min_{u, v \in X', u \neq v} \dist(u, v)}.\) The following lemma by Gupta et al.~\cite{1238226} shows that the cardinality of a subset~\(X' \subseteq X\) can be bounded by its aspect ratio and~\(\dd\).
\begin{lemma}[\cite{1238226}]
  \label{lem:upper-bound-on-number-of-vertices-low-dd}
  Let~\((X, \dist)\) be a metric and \(\dd\)~its doubling dimension.
  Consider a subset~\({X' \subseteq X}\) of aspect ratio~\(\alpha\) and doubling dimension~\(\dd'\).
  Then it holds that~\(\dd' = \bigO(\dd)\) and~\(\abs{X'} \leq 2^{\bigO(\dd \lceil \log_2 \alpha \rceil)}\).
\end{lemma}

Using Lemma~\ref{lem:upper-bound-on-number-of-vertices-low-dd}, we bound the size of~\(Y\).

\begin{lemma}
  \label{lem:ckswo-delta-net-upper-bound}
  Let~\(\mathcal{I} = (G, k, p, L)\) be an instance of the \problemName{CkSwO} problem, \({\varepsilon > 0}\), and \({\varrho \in \R^+}\)~a cost.
  Moreover let~\((\Vsuppliers, \dist)\) be the shortest-path metric induced by~\(\Vsuppliers\) and \(\dd\)~its doubling dimension.
  Assume that for each supplier~\(s \in \Vsuppliers\) there exists a client~\(c \in \Vclients\) such that~\(\dist(s, c) \leq \varrho\).
  If~\(\mathcal{I}\) has a feasible solution~\(\phi\) with~\(\cost(\phi) \leq \varrho\), then an~\((\varepsilon\varrho)\)-net~\(Y\) of~\(\Vsuppliers\) has size at most~\((k + p)\varepsilon^{-\bigO(\dd)}\).
\end{lemma}

Before we prove the lemma, let us make a few comments the statement of the lemma.
We do not know the cost of the optimum solution and we are merely guessing it.
Hence we need to also consider the case when our guess on the cost~\(\varrho\) is wrong, i.e.~it is less than the cost of the optimum solution.
The requirement that every supplier has a client nearby is a natural one:
if we assume that our solution has cost~\(\varrho\) and a supplier~\(s\) has~\(\dist(s, \Vclients) > \varrho\), then it will never be picked in a solution.
Thus we can without loss of generality remove all such suppliers from the input.

\begin{proof}[of {Lemma~\ref{lem:ckswo-delta-net-upper-bound}}]
  Let~\(\Vclients' = \{u \in \Vclients \mid \phi(u) \neq \outlier\}\) and~\(\Vsuppliers' = \phi(\Vclients')\).
  Since~\(\phi\) is a solution of cost~\(\varrho\), all clients of~\(\Vclients'\) can be covered by balls of radius~\(\varrho\) around vertices of~\(\Vsuppliers'\) and there are~\(k\)~such balls.
  To cover the outliers~\(\Vclients \setminus \Vclients'\), we use the fact that there are at most~\(p\) of them.
  Thus we place a ball of radius~\(\varrho\) centered at each outlier.
  Since every supplier is at distance at most~\(\varrho\) from some client and every client lies in one of the balls of radius~\(\varrho\) we have placed, by increasing the radius of every placed ball to~\(2\varrho\) we cover all suppliers as well.

  In total, \(\Vclients \cup \Vsuppliers\) can be covered by~\(k + p\)~balls of diameter at most~\(4\varrho\).
  For each such ball, let us consider the aspect ratio of points of the net~\(Y\) which lie in it.
  Since the diameter is at most~\(4\varrho\) and the distance of every two points of a net is more than~\(\varepsilon\varrho\), the aspect ratio is at most~\(\frac{4}{\varepsilon}\).
  By Lemma~\ref{lem:upper-bound-on-number-of-vertices-low-dd} each of the balls contains at most~\(\varepsilon^{-\bigO(\dd)}\) points of~\(Y\).
  Thus~\(\abs{Y} \leq (k + p)\varepsilon^{-\bigO(\dd)}\).
  \qed
\end{proof}

When we gave the intuition behind the algorithm, we assumed that the derived solution~\(\phi\), which replaces every optimum supplier of~\(\phi^*(\Vclients) \setminus \{\outlier\}\) by its nearest net point, does not violate the capacity of any selected net point, i.e.~for every~\(s \in \phi(\Vclients) \setminus \{\outlier\}\) we have~\(\abs{\phi^{-1}(s)} \leq L(s)\).
This does not have to be the case, so instead of replacing every optimum supplier by its nearest net point, we need to select the replacement net point in a more sophisticated manner, in particular to avoid violating the capacity of the replacement net point.

Let~\(\Vsuppliers^*\) be the optimum supplier set corresponding to the optimum assignment function~\(\phi^*\).
Suppose that we are able to guess a subset~\(S^* \subseteq Y\) of size~\(k\) such that for every supplier~\(u \in \Vsuppliers^*\) we have~\(\dist(u, S^*) \leq \varepsilon\varrho\).
Let~\(A \colon \Vsuppliers^* \to S^*\) map each optimum supplier to its nearest net point.
As we have discussed, we cannot just replace each supplier~\(u \in \Vsuppliers^*\) by~\(A(u)\) since it may happen that~\({\abs{(\phi^*)^{-1}(u)} > L(A(u))}\).
However, there is a supplier in the ball~\(\ball{A(u)}{\varepsilon\varrho}\) which is guaranteed to have capacity at least~\(L(u)\) since~\(u \in \ball{A(u)}{\varepsilon\varrho}\).
Thus we can implement the ``replacement step'' by replacing each optimum supplier by the supplier of highest capacity in~\(\ball{A(u)}{\varepsilon\varrho}\) and this increases the cost of the optimum solution by at most~\(2\varepsilon\varrho\), i.e.~the diameter of the ball.

We must also consider the case when~\(\abs{A^{-1}(v)} > 1\) for some net point~\(v \in S^*\).
Generalizing the previous idea, we replace suppliers~\(A^{-1}(v)\) by~\(\abs{A^{-1}(v)}\)~suppliers of~\(\ball{v}{\varepsilon\varrho}\) with the highest capacities.
As we do not know the optimum solution, we do not know~\(\abs{A^{-1}(v)}\) either.
Nevertheless, we know that~\(\abs{S^*} \leq k\), and so we can afford to guess these values after guessing the set~\(S^*\).

\begin{figure}[tb]
  \centering
  \begin{subfigure}{.48\textwidth}
    \centering
    \includegraphics{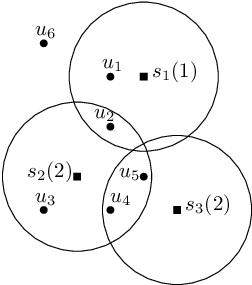}
  \end{subfigure}
  \begin{subfigure}{.48\textwidth}
    \centering
    \includegraphics{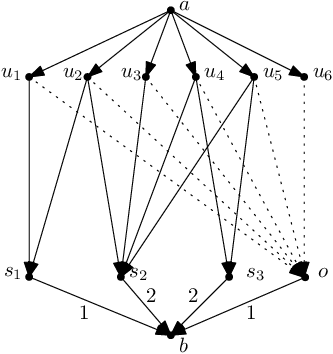}
  \end{subfigure}
  \caption{%
    On the left, we have a \problemName{CkSwO} instance with~one~allowed outlier.
    Vertices~\(s_1\), \(s_2\), and~\(s_3\) are suppliers and circles centered at them have radius~\(\varrho\).
    The remaining vertices are clients.
    The capacities are~\(L(s_1) = 1\), \(L(s_2) = 2\), \(L(s_3) = 2\).
    On the right, we have the result of applying Lemma~\ref{lem:recognize-feasible-ckswo-solution} on the instance in the left figure.
    Edges with unspecified capacity have capacity~\(1\).
  }
  \label{fig:ckswo-feasibility}
\end{figure}

The final ingredient we need is the ability to verify our guesses.
That is, given a set of at most~\(k\)~suppliers, we need to check if there exists a feasible solution of a given cost which assigns clients to a prescribed set of suppliers.
We do so by a standard reduction to network flows.
\begin{lemma}
  \label{lem:recognize-feasible-ckswo-solution}
  Given a \problemName{CkSwO} instance~\(\mathcal{I} = (G, k, p, L)\), a cost~\(\varrho \in \R^+\), and a subset~\(S \subseteq \Vsuppliers\), we can determine in polynomial time whether there exists an assignment~\(\phi \colon \Vclients \to S\) such that~\(\abs{\phi(u)^{-1}} \leq L(u)\) for each~\(u \in S\), \(\abs{\phi^{-1}(\outlier)} \leq p\), and~\(\cost(\phi) \leq \varrho\).
\end{lemma}
\begin{proof}
  We build a network~\(N = (G' = (V', E'), a, b, c)\) where~\(a \in V'\) is the source, \(b \in V'\) is the sink, and~\(c: E' \to \R^+_0\) assigns capacities to edges of the directed graph~\(G' = (V', E')\).
  Refer to Fig.~\ref{fig:ckswo-feasibility} for an example of such a network.
  For each vertex of~\(S \cup \Vclients\) we are going to create a unique vertex in~\(V'\), we denote this bijection by~\(m\).
  For a subset~\(U \subseteq S \cup \Vclients\) we denote~\(m(U) = \{m(u) \mid u \in U\}\).
  For each client~\(u \in \Vclients\) we add a vertex~\(v_u\) to~\(V'\), assign~\(m(u) = v_u\), and add an edge from~\(a\) to~\(v_u\) with capacity~\(1\).
  For each supplier~\(s \in S\) we add a vertex~\(v_s\) to~\(V'\) and assign~\(m(s) = v_s\).
  We add edges from each vertex of~\(\{m(u) \mid u \in \Vclients \cap \ball{s}{\varrho}\}\) to~\(v_s\) with capacity~\(1\).
  We add an edge from~\(v_s\) to~\(b\) with capacity~\(L(s)\).
  Finally we add a vertex~\(o\) and add edges from each~\(m(\Vclients)\) to~\(o\) with capacity~\(1\) and an edge from~\(o\) to~\(b\) with capacity~\(p\).

  We claim that the desired assignment~\(\phi: \Vclients' \to S\) exists if and only if the maximum flow of~\(N\) is~\(\abs{\Vclients}\).

  To prove the forward implication, let~\(\phi\) be a feasible solution of the instance~\(\mathcal{I}\).
  For each client~\(u\) such that~\(\phi(u) \neq \outlier\), we send a unit flow on the path~\((a, m(u), m(\phi(u)), b)\).
  As~\(\phi\) is a feasible solution, for each~\(s \in \Vsuppliers\) there are at most~\(L(s)\) units flowing through the edge~\((m(s), b)\).
  For each outlier~\(v \in \phi^{-1}(\outlier)\) we send a unit of flow on the path~\((a, m(v), o, b)\).
  The solution~\(\phi\) creates at most~\(p\) outliers, thus the flow through the edge~\((o, b)\) is at most~\(p\), which does not exceed the capacity of the edge~\((o, b)\).
  There is one unit of flow going through each client, hence the flow has value~\(\abs{\Vclients}\).
  All edges leaving the source~\(a\) are saturated and there are no incoming edges to~\(a\) hence this flow is also maximum.

  To prove the backward implication, let the maximum flow of the network~\(N\) be~\(\abs{\Vclients}\) and let~\(f\) be the corresponding flow.
  Assume that~\(f\) assigns an integral flow to each edge, this is without loss of generality since we can compute such a flow using the Edmonds--Karp~\cite{edmonds1972theoretical} algorithm, and it is a well known fact that if the capacities are integral, then the flow through each edge will be integral as well.
  To simplify the exposition, let~\(E' = E' \setminus \{e \in E' \mid f(e) = 0\}\), that is we remove all edges through which there is no flow.
  From the construction of network~\(N\) and the preceding assumptions, each vertex of~\(m(\Vclients)\) has one outgoing edge and this edge has its other endpoint in~\(m(S) \cup \{o\}\).
  We construct a feasible solution~\(\phi\) to the instance~\(\mathcal{I}\) as follows.
  If the outgoing edge from~\(m(u)\) of a client~\(u \in \Vclients\) ends in~\(o\), then we set~\(\phi(u) = \outlier\).
  Otherwise the outgoing edge from~\(m(u)\) ends in~\(m(s)\) of some supplier~\(s \in \Vsuppliers\), then we set~\(\phi(u) = s\).
  Since the capacity of the edge~\((o, b)\) is~\(p\) there can be at most~\(p\) outliers in our solution~\(\phi\).
  From the construction of the network, the distance between a client and its assigned supplier is at most~\(\varrho\).
  Finally, the capacity of the edge~\((m(s), b)\) for a supplier~\(s \in S\) is set to~\(L(s)\), hence there cannot be more than~\(L(s)\) clients assigned to~\(s\).

  Clearly, the construction of the network~\(N\) can be done in polynomial time.
  The aforementioned Edmonds--Karp algorithm~\cite{edmonds1972theoretical} has a polynomial running time.
  Note that we have also described how to compute a feasible solution to~\(\mathcal{I}\) of cost~\(\varrho\) such that~\(S\) is the set of opened suppliers from a maximal flow in~\(N\).
  \qed
\end{proof}

We now prove the correctness of the replacement strategy.

\begin{lemma}
  \label{lem:ckswo-epas-dd-main}
  Let~\(\mathcal{I} = (G, k, p, L)\) be a \problemName{CkSwO} instance such that there exists a solution~\(\phi^*\) of cost~\(\varrho\) and for each supplier there exists a client at distance at most~\(\varrho\) from it.
  Given an~\((\varepsilon\varrho)\)-net~\(Y\) of the shortest-path metric induced by~\(\Vsuppliers\), and~\(\varepsilon > 0\), we can compute a solution of cost~\((1 + 2\varepsilon)\varrho\) in time~\(\bigO^*\left(\binom{\abs{Y}}{k} k^k\right)\).
\end{lemma}
\begin{proof}
  \label{lem:compute-ckswo-sol-from-delta-net}
  Let~\(\Vsuppliers^* = \phi^*(\Vclients) \setminus \{\outlier\}\).
  For an optimum supplier~\(u \in \Vsuppliers^*\) it may happen that~\(\abs{\ball{u}{\varepsilon\varrho} \cap Y} > 1\), i.e.~it is close to more than one net point.
  This may cause issues when we guess for each net point~\(v \in Y\) the size of~\(\ball{v}{\varepsilon\varrho} \cap \Vsuppliers^*\).
  To circumvent this problem, we fix a linear order~\(\preceq\) on the set of net points~\(Y\) and we assign each optimum supplier to the first close net point.
  Formally, we define for a net point~\(v \in Y\)
  \begin{itemize}
    \item \(P(v) = \{v' \in Y \mid v' \prec v\}\) (note that~\(v \not\in P(v)\)),
    \item \(M(v) = \ball{v}{\varepsilon\varrho} \setminus (\cup_{v' \in P(v)} \ball{v'}{\varepsilon\varrho})\),
    \item \(D(v) = \abs{M(v) \cap \Vsuppliers^*}\), and
    \item \(R(v)\) to be the set of~\(D(v)\) suppliers in~\(M(v)\) with the highest capacities.
  \end{itemize}
  For a net point~\(v \in Y\), it is easy to see that~\(\sum_{s \in R(v)} L(s) \geq \sum_{t \in M(v) \cap \Vsuppliers^*} L(t)\).
  The sets~\(\{R(v) \mid v \in Y\}\) are disjoint by the way we defined~\(M(v)\).

  We guess a subset~\(Y' \subseteq Y\) of size~\(k\) such that~\(\Vsuppliers^* \subseteq \bigcup_{v \in Y'} \ball{v}{\varepsilon\varrho}\).
  For each~\(v \in Y'\) we guess~\(D(v)\) and select~\(S = \bigcup_{v \in Y'} R(v)\).
  We apply the algorithm from Lemma~\ref{lem:recognize-feasible-ckswo-solution} with the set~\(S\) and cost~\((1 + 2\varepsilon)\varrho\).
  If this check passes, then the solution we obtain is in fact a solution of cost~\((1 + 2\varepsilon)\varrho\) since we replaced each optimum supplier by a supplier at distance at most~\(2\varepsilon\varrho\) from it.
  Conversely, if none of our guesses pass this check, then the instance~\(\mathcal{I}\) has no solution of cost~\(\varrho\).

  The running time of our algorithm is dominated by the time required to guess~\(Y'\) and the cardinalities~\(D(v)\) for each~\(v \in Y'\).
  From~\(\abs{Y'} \leq k\), the time required to guess~\(Y'\) is~\(\bigO\left(\binom{\abs{Y}}{k}\right)\).
  Since~\(D(v) \leq k\) for every~\(v \in Y\), the time required to guess~\(D(v)\) for each~\(v \in Y'\) is~\(\bigO(k^k)\).
  In total, the running time of the algorithm is~\(\bigO^*\left(\binom{\abs{Y}}{k} k^k \right)\).
  \qed
\end{proof}

We are ready to prove Lemma~\ref{lem:ckswo-epas-dd}.
\begin{proof}[of {Lemma~\ref{lem:ckswo-epas-dd}}]
  To satisfy the requirements of Lemma~\ref{lem:ckswo-delta-net-upper-bound} and Lemma~\ref{lem:ckswo-epas-dd-main}, we remove all suppliers which do not have a client at distance~\(\varrho\) from them.
  The algorithm then computes a~\((\varepsilon\varrho)\)-net~\(Y\) of the metric~\((\Vsuppliers, \dist)\).
  By Lemma~\ref{lem:ckswo-delta-net-upper-bound}, if the net~\(Y\) has more than~\((k + p)\varepsilon^{-\bigO(\dd)}\) points, then the algorithm concludes that the instance has no solution of cost~\(\varrho\).
  To compute a solution of cost~\((1 + 2\varepsilon)\varrho\) or to show that there exists no solution of cost~\(\varrho\), we apply the algorithm given by Lemma~\ref{lem:ckswo-epas-dd-main}.
  Since the net~\(Y\) has size at most~\((k + p)\varepsilon^{-\bigO(\dd)}\), the running time of the algorithm is~\(\bigO^*\left((k + p)^k\varepsilon^{-\bigO(k\dd)}\right)\).
  \qed
\end{proof}

\section{EPAS for \problemName{kSwO} in Low Highway Dimension Graphs}
\label{sec:hd-kswo}

In this section we prove Theorem~\ref{thm:kswo-epas-hd}, i.e.~we show an EPAS for \problemName{\(k\)\nobreakdash-Supplier with Outliers} in low highway dimension graphs.
We assume without loss of generality that edge lengths are integral.
In our algorithm, we will require the constant~\(\gamma\) in Definition~\ref{defi:hd1} to be strictly greater than~\(4\).

In contrast to \problemName{CkSwO}, we can simply specify the solution of \problemName{kSwO} for a given cost~\(\varrho\) by a subset of suppliers~\(S \subseteq \Vsuppliers\).
Then a client~\(c\) can be assigned to any supplier~\(s\) such that~\(\dist(c, s) \leq \varrho\); if such a supplier does not exist, then~\(c\) is an outlier.
For such a solution~\(S\), we denote the set of outliers by~\(S^\outlier\).

Let us first give an overview of our algorithm.
We use a framework by Becker et al.~\cite{becker_et_al:LIPIcs:2018:9471} for embedding low highway dimension graphs into low treewidth graphs.
An \emph{embedding} of an (undirected) \emph{guest graph}~\(G\) into a \emph{host graph}~\(H\) is a distance\nobreakdash-preserving injective mapping~\(\psi \colon V(G) \to V(H)\).
For the purpose of this overview, suppose that for every two vertices~\(u, v \in V(G)\) we have~\(\dist_G(u, v) \leq \dist_H(\psi(u), \psi(v)) \leq c \cdot \dist_G(u, v)\) for some constant~\(c \in \R^+\).
Ideally, we would like to solve the problem optimally in the host graph and then translate the solution in the host graph back into the input guest graph.
The resulting solution in the guest graph would be a~\(c\)\nobreakdash-approximate solution of the input instance.

Our situation is complicated by the fact that optimally solving \problemName{\(k\)\nobreakdash-Center}, which is a special case of \problemName{kSwO}, is already \(\classW{1}\)\nobreakdash-hard in low treewidth graphs, as was shown by Katsikarelis et al.~\cite{KATSIKARELIS201990}.
On the bright side, the authors of the same paper show that it is possible to design an EPAS for \problemName{\(k\)\nobreakdash-Center} in low treewidth graphs.
We generalize their algorithm to \problemName{kSwO}.
The following theorem summarizes the properties of the algorithm we obtain.
\begin{theorem}
  \label{thm:kswo-epas-tw}
  Let~\({\mathcal{I} = (G, k, p)}\) be a \problemName{kSwO} instance, \(\varrho \in \N\) a cost, and~\({\varepsilon > 0}\).
  Suppose that we receive a nice tree decomposition\footnote{See Section~\ref{sec:kswo-epas-tw} for a formal definition.} of~\(G\) of width~\(\tw(G)\) on input.
  There exists an algorithm which either
  \begin{itemize}
    \item returns a solution of cost~\((1 + \varepsilon)\varrho\) if~\((1 + \varepsilon)\varrho \geq \OPT(\mathcal{I})\), or
    \item correctly decides that~\(\mathcal{I}\) does not have a feasible solution of cost~\(\varrho\),
  \end{itemize}
  running in time~\(\bigO^*\left((\tw(G) / \varepsilon)^{\bigO(\tw(G))}\right)\).
\end{theorem}

To ease the presentation, we defer the full proof of Theorem~\ref{thm:kswo-epas-tw} to Section~\ref{sec:kswo-epas-tw}, however, let us at least give the idea behind the algorithm.
We use the standard approach of dynamic programming on nice tree decompositions.
We use a technique of Lampis~\cite{10.1007/978-3-662-43948-7_64} for designing approximation schemes for problems hard in low treewidth graphs.
The idea behind his technique is that if the cause of the hardness of a problem is due to the necessity of storing large numbers in the dynamic programming table, which in our case are integers from~0 to~\(\varrho\), then instead of storing exact values we store powers of~\((1 + \delta)\) for an appropriately selected~\(\delta\).
This decreases the size of the dynamic programming table from~\(\varrho^\tw\) to roughly~\((\log \varrho)^{\bigO(\tw)}\), which is sufficient for us.
This approach is not cost\nobreakdash-free though, since storing rounded approximate values may create errors which can accumulate during the execution of the algorithm.
We will show how to bound this error in the height of the given tree decomposition.
Together with a result of Chatterjee et al.~\cite{chatterjee2014optimal}, which rebalances a given tree decomposition into a nice tree decomposition with a logarithmic height and while only increasing the width by a constant factor, we will obtain an EPAS parameterized by treewidth of the input graph.

The main result by Becker et al.~\cite{becker_et_al:LIPIcs:2018:9471} that we use is the following theorem.
Note that this result requires that the universal constant~\(\gamma\) in Definition~\ref{defi:hd1} is strictly greater than~\(\gamma\).
\begin{theorem}[{\cite[Theorem~4]{becker_et_al:LIPIcs:2018:9471}}]
  \label{thm:becker-embedding}
  There is a function~\(f(\cdot, \cdot, \cdot)\) such that, for every~\(\varepsilon > 0\), graph~\(G\) of highway dimension~\(\hd\), and set of vertices~\(U \subseteq V(G)\), there exists a graph~\(H\) and an embedding~\(\psi \colon V(G) \to V(H)\) such that
  \begin{itemize}
    \item \(H\) has treewidth at most~\(f(h, \abs{U}, \varepsilon)\) and
    \item for all vertices~\(u, v \in V(G)\)
      \begin{equation}
        \label{eqn:becker-upper-bound}
        \begin{split}
          \dist_G(u, v) & \leq \dist_H(\psi(u), \psi(v)) \\
                        & \leq (1 + \bigO(\varepsilon))\dist_G(u, v) + \varepsilon \cdot \min\{\dist_G(u, U), \dist_G(v, U)\}.
        \end{split}
      \end{equation}
  \end{itemize}
\end{theorem}

Our algorithm starts by computing a constant factor approximation solution~\(S\) to the input instance, we can compute such solution using the result of Cygan and Kociumaka~\cite{cygan_et_al:LIPIcs:2014:4462}.
We select~\(S \cup S^\outlier\) as the set~\(U\) in Theorem~\ref{thm:becker-embedding}.
Since~\(S\) is a feasible solution, we have~\(\abs{U} \leq k + p\).
We apply the algorithm of Theorem~\ref{thm:becker-embedding} on graph~\(G\) to obtain a host graph~\(H\) with an embedding~\(\psi \colon V(G) \to V(H)\).
For a subset of suppliers~\(S \subseteq \Vsuppliers\) let~\(\cost_G(S)\) be the cost~\(S\) in graph~\(G\) with its edge lengths.

Let~\(\OPT_G\) and~\(\OPT_H\) be the optimum cost of the instance~\(G\) and~\(H\) respectively.
We apply the algorithm of Theorem~\ref{thm:kswo-epas-tw} on graph~\(H\) with values~\(\varrho\) and~\(\varepsilon\) to obtain a solution~\(S\) such that~\(\cost_H(S) \leq (1 + \varepsilon)\varrho\).
As~\(\psi\) is an injective function, it is correct to refer to~\(S\) as a solution in~\(G\).
Our goal is to bound the cost of the obtained solution in the input graph~\(G\), which we do in the following lemma.
\begin{lemma}
  \label{lem:kswo-epas-hd-cost}
  Let~\(S\) be a~\((1 + \varepsilon)\)\nobreakdash-approximate solution in the host graph~\(H\), i.e.~\({\cost_H(S) \leq (1 + \varepsilon)\OPT_H}\).
  Then~\(\cost_G(S) \leq (1 + \bigO(\varepsilon))\OPT_G\).
\end{lemma}
\begin{proof}
  Let~\(S^*\) be a solution of~\(G\) of cost~\(\OPT_G\) and~\(\Vclients^* = (\Vclients \setminus (S^*)^\outlier)\).
  For a client~\(u \in \Vclients^*\) we denote by~\(S^*(u)\) the nearest supplier of~\(S^*\) to~\(u\).
  By definition of~\problemName{kSwO} we have~\(\cost_H(S^*) = \max_{u \in \Vclients^*} \dist_H(u, S)\).
  From~\eqref{eqn:becker-upper-bound} we get
  \begin{multline}
    \label{eqn:kswo-becker-ub-from-defi}
    \cost_H(S^*) \leq \\
    \max_{u \in \Vclients^*}\{%
      (1 + \bigO(\varepsilon))\dist_G(u, S^*(u))
      + \varepsilon \cdot \min\{\dist_G(u, U), \dist_G(S^*(u), U)\}
    \}.
  \end{multline}

  By maximizing over each term of~\eqref{eqn:kswo-becker-ub-from-defi} and using the fact that~\({\min\{a, b\} \leq a}\) for any~\({a, b \in \R^+_0}\) in the second term we get 
  \begin{equation}
    \label{eqn:kswo-becker-before-using-outliers}
    \cost_H(S^*)
    \leq 
    (1 + \bigO(\varepsilon))\OPT_G + \varepsilon \cdot \max_{u \in \Vclients^*} \dist_G(u, U).
  \end{equation}
  We are going to give an upper bound for the second term of~\eqref{eqn:kswo-becker-before-using-outliers} using the first term.
  There are two cases to consider.
  If~\(u\) is covered by the constant-approximate solution, then~\(\dist_G(u, U) = \bigO(\OPT_G)\).
  Otherwise~\(u\) has to be an outlier in the approximate solution.
  We added all outliers to the set~\(u\), so we have~\({\dist_G(u, U) = 0}\) in this case.
  In total, we have
  \begin{equation}
    \label{eqn:kswo-becker-bound-apx-sol}
    \cost_H(S^*) \leq (1 + \bigO(\varepsilon))\OPT_G.
  \end{equation}
  As~\(S^*\) is an optimal solution in~\(G\) but not necessarily in~\(H\) we get from~\eqref{eqn:kswo-becker-bound-apx-sol}
  \begin{equation}
    \label{eqn:kswo-becker-last-step}
    \OPT_H \leq \cost_H(S^*) \leq  (1 + \bigO(\varepsilon))\OPT_G
  \end{equation}
  The approximate solution has cost~\((1 + \varepsilon)\OPT_H\).
  By multiplying both sides of~\eqref{eqn:kswo-becker-last-step}, we get the desired bound.
  \qed
\end{proof}

We are ready to prove Theorem~\ref{thm:kswo-epas-hd}.
\begin{proof}[of Theorem~\ref{thm:kswo-epas-hd}.]
  Lemma~\ref{lem:kswo-epas-hd-cost} shows the correctness of the algorithm, it remains to bound its running time.
  The algorithm consists of three steps, first we compute a constant\nobreakdash-factor approximation to the instance in polynomial time.
  Then we run the algorithm from Theorem~\ref{thm:becker-embedding} on the input to obtain a host graph~\(H\).
  This step runs in time~\(\bigO^*(g(k, p, \hd, \varepsilon))\) for some computable function~\(g\).
  Finally, we run the algorithm from Theorem~\ref{thm:kswo-epas-tw} on~\(H\).
  This step takes time~\(\bigO((\tw(H) / \varepsilon)^{\bigO(\tw(H))})\) where~\(\tw(H) = \bigO^*(g'(k, p, \hd, \varepsilon))\) for some computable function~\(g'\).
  Thus the algorithm runs in~\(\classFPT\) time with parameters~\(k\), \(p\), \(\varepsilon\), and the highway dimension of the input.
\end{proof}

\section{EPAS for \(k\)\nobreakdash-Supplier with Outliers on Low Treewidth Graphs}
\label{sec:kswo-epas-tw}

The following definition of treewidth is given in~\cite{10.1007/978-3-319-21275-3}.
A~\emph{tree decomposition} of a graph~\(G\) is a pair~\(\mathcal{T} = (T, \{X_t\}_{t \in V(T)})\), where~\(T\) is a tree whose every node~\(t\) is assigned a vertex subset~\(X_t \subseteq V(G)\), called a \emph{bag}, such that the following three~properties hold.
\begin{property}
  Every vertex of~\(G\) is in at least one bag, i.e.~\(\cup_{t \in V(T)} X_t = V(G)\).
\end{property}
\begin{property}
  \label{defi:tree-decomposition--prop-2}
  For every edge~\(e \in E(G)\), there exists a node~\(t \in T\) such that the corresponding bag~\(X_t\) contains both endpoints of~\(e\).
\end{property}
\begin{property}
  \label{defi:tree-decomposition--prop-3}
  For every~\(v \in V(G)\), the set~\(T_u = \{t \in V(T) \mid u \in X_t\}\), that is the set of nodes whose corresponding bags contain~\(u\), induces a connected subtree of~\(T\).
\end{property}
To improve comprehensibility, we shall refer to vertices of the underlying tree as \emph{nodes}.
It follows from Property~\ref{defi:tree-decomposition--prop-3} that for all~\(i, j, k \in V(T)\), if~\(j\) is on the path from~\(i\) to~\(k\) in~\(T\), then~\(X_i \cap X_k \subseteq X_j\).
The \emph{width} of a tree decomposition~\(\mathcal{T} = (T, \{X_t\}_{t \in V(T)})\) equals~\(\max_{t \in V(T)} (\abs{X_t} - 1)\), that is, the maximum size of its bag minus~\(1\).
The \emph{treewidth} of a graph~\(G\), denoted by~\(\tw(G)\), is the minimum possible width of a tree decomposition of~\(G\).

For algorithmic purposes, it is often more convenient to work with nice tree decompositions:
a rooted tree decomposition~\((T, \{X_t\}_{t \in V(T)})\) with root~\(r \in V(T)\) is a \emph{nice tree decomposition} if each of its leaves~\(\ell \in V(T)\) contains an empty bag (that is~\(X_\ell = \emptyset\)) and inner nodes are one of the following three types:
\begin{itemize}
  \item \textbf{Introduce node:} a node~\(t\) with exactly one~child~\(t'\) where~\(X_t = X_{t'} \cup \{u\}\) for some vertex~\(u \not\in X_{t'}\). We say that~\(u\) is \emph{introduced} at~\(t\).
  \item \textbf{Forget node:} a node~\(t\) with exactly one~child~\(t'\) where~\(X_t = X_{t'} \setminus \{v\}\) for some vertex~\(v \in X_{t'}\). We say that~\(v\) is \emph{forgotten} at~\(t\).
  \item \textbf{Join node:} a node~\(t\) with exactly two~children~\(t_1, t_2\) where~\(X_t = X_{t_1} = X_{t_2}\).
\end{itemize}
It is known that if graph~\(G\) admits a tree decomposition of width at most~\(k\), then it also admits a nice tree decomposition of width at most~\(k\).
Furthermore, given a tree decomposition~\(\mathcal{T} = (T, \{X_t\}_{t \in V(T)})\) of~\(G\) of width at most~\(k\), one can in~\(\bigO(k^2 \cdot \max(\abs{V(T)}, \abs{V(G)}))\)~time compute a nice tree decomposition of~\(G\) of width at most~\(k\) that has at most~\(\bigO(k \cdot \abs{V(G)})\)~nodes, for more details, see~\cite[Lemma 7.4]{10.1007/978-3-319-21275-3}.
For this reason we shall always assume without loss of generality that input tree decompositions of our algorithms are nice.
By~\(V_t\) we denote vertices which appear in bags in the subtree rooted at the vertex corresponding to~\(X_t\) and by~\(G[X_t]\) and~\(G[V_t]\) we denote the subgraph induced by vertices in bag~\(X_t\) and vertices~\(V_t\) respectively.

In this section we develop an EPAS for \problemName{kSwO} parameterized by~\(k\), \(p\), \(\varepsilon\), and the treewidth of the input graph to prove Theorem~\ref{thm:kswo-epas-tw}.
To simplify the exposition, we restrict ourselves to instances with integral edge lengths, and we also assume that~\(\Vsuppliers \cap \Vclients = \emptyset\).
We start by giving an exact algorithm for the problem, which we later turn into an approximation scheme by the technique of Lampis~\cite{10.1007/978-3-662-43948-7_64}.
The overall approach generalizes the approach for obtaining an EPAS for \problemName{\(k\)\nobreakdash-Center} by Katsikarelis et al.~\cite{KATSIKARELIS201990}.
In the rest of this section, we denote the treewidth of the input graph by~\(\tw\).
The properties of the exact algorithm are summarized by the following theorem.
\begin{theorem}
  There exists an algorithm which given a \problemName{kSwO} instance~\(\mathcal{I}\) and cost~\(\varrho \in \N\) decides whether~\(\mathcal{I}\) has a feasible solution of cost~\(\varrho\) running in time~\(\bigO^*\left(\varrho^{\bigO(\tw)}\right)\).
  \label{thm:kswo-xp-alg-tw}
\end{theorem}

We give an equivalent formulation of~\problemName{kSwO} which is more convenient for our purposes.
Let~\((G, k, p)\) be an instance of the~\problemName{kSwO} problem where~\(G = (V, E)\) has edge lengths~\(d \colon E \to \N\) and vertices are partitioned into clients~\(\Vclients\) and suppliers~\(\Vsuppliers\).
A \emph{distance labelling} function~\(\dl\) of~\(G\) is a function~\(\dl \colon V \to \{0, \ldots, \varrho\} \cup \{\infty\}\).
We require that only suppliers can have label~\(0\).
We say that a vertex~\(u \in V\) is \emph{satisfied} by~\(\dl\) if~\(\dl(u) = 0\) or~\(u\) has a finite label and there exists a neighbor~\(v \in N(u)\) such that~\(\dl(u) \geq \dl(v) + d(u, v)\).
Given a distance labelling function~\(\dl\), if every client is either satisfied or has label~\(\infty\), then we say that~\(\dl\) is \emph{valid}.
We define the \emph{cost} of a distance labelling function~\(\dl\) as~\(\abs{\dl^{-1}(0)}\) and the \emph{penalty} as~\(\abs{\dl^{-1}(\infty) \cap \Vclients}\).
The following lemma shows the equivalence between the two formulations.
A special case of this statement for~\problemName{\(k\)\nobreakdash-Center} is proved in the original algorithm, cf.~\cite[Lemma~26]{KATSIKARELIS201990}.
\begin{lemma}
  \label{lem:kswo-epas-tw--dl}
  A \problemName{kSwO} instance~\(\mathcal{I} = (G, k, p)\) admits a feasible solution of cost~\(\varrho\) if and only if it admits a valid distance labelling function~\(\dl \colon V \to \{0, \ldots, \varrho\} \cup \{\infty\}\) of cost~\(k\) and penalty~\(p\).
\end{lemma}
\begin{proof}
  Let~\(S \subseteq \Vsuppliers\) be a solution of~\(\mathcal{I}\) of cost~\(\varrho\).
  We construct the required distance labelling function~\(\dl\) as follows:
  \begin{enumerate}
    \item for each vertex~\(u \in V\) with~\(\dist(u, S) \leq \varrho\) we set~\(\dl(u) = \dist(u, S)\),
    \item we set the labels of the remaining vertices to~\(\infty\).
  \end{enumerate}
  It is immediate that the cost of~\(\dl\) is at most~\(k\).
  From the definition of~\problemName{kSwO} it is also clear that outliers are clients whose distance from~\(S\) exceeds~\(\varrho\), therefore the penalty of~\(\dl\) is at most~\(p\).
  Consider a vertex~\(u \in V\) and a shortest path~\(\pi\) of length at most~\(\varrho\) between~\(u\) and its nearest opened supplier~\(s \in S\).
  Let~\(v\) be the neighbor of~\(u\) on~\(\pi\).
  We have~\(\dist(u, s) = d(u, v) + \dist(v, s)\) and~\(\dl(u) = \dist(u, s)\) and~\(\dl(v) = \dist(v, s)\), this shows that every vertex and particularly every client with a finite label is satisfied.

  For the opposite direction, let~\(\dl\) be a distance labelling function with properties given by the lemma statement.
  We select vertices with label~\(0\) as the solution~\(S\).
  By definition of a distance labelling function these can only be suppliers.
  It suffices to show that the distance of vertices with finite labels from~\(S\) is at most the value of their label.
  This is the case because a valid distance labelling function assigns finite labels to all but (up to) \(p\)~vertices and the maximum finite label is~\(\varrho\).
  We prove this claim by induction on label values.
  The base case is immediate since vertices with label~\(0\) are precisely the solution~\(S\).
  In the induction step consider a vertex~\(u\) with a finite label~\(\ell\).
  Since~\(u\) is a satisfied vertex, there exists a neighbor~\(v \in N(u)\) such that~\(\dl(u) \geq \dl(v) + d(u, v)\).
  As edge lengths are positive, we have~\(\dl(u) > \dl(v)\) and using triangle inequality we obtain
  \begin{equation}
    \dist(u, S) \leq \dist(u, v) + \dist(v, S) \leq \dist(u, v) + \dl(v) \leq \dl(u)
  \end{equation}
  where the second inequality follows from the induction hypothesis~\(\dist(v, S) \leq \dl(v)\).
  This shows that clients with finite labels are covered by the solution~\(S\) and since the penalty of~\(\dl\) is at most~\(p\), there are at most~\(p\) outliers.
  Also the cost of~\(\dl\) is at most~\(k\) thus we also have~\(\abs{S} \leq k\).
  \qed
\end{proof}

According to the proven equivalence, we may refer to a client with label~\(\infty\) as an outlier and a supplier with label~\(0\) as an opened supplier.

We are ready to prove Theorem~\ref{thm:kswo-xp-alg-tw}.
The algorithm will be a standard dynamic programming procedure on a nice tree decomposition of the input graph.
As is customary, let us assume that a nice tree decomposition of the input graph~\(G\) is given as a part of the input.
\begin{proof}[{Theorem~\ref{thm:kswo-xp-alg-tw}}]
  Let~\(\mathcal{T} = (T, \{X_t\}_{t \in V(T)})\) be a nice tree decomposition of the input graph~\(G\) of width~\(\tw(G)\).
  For every node of~\(i \in V(\mathcal{T})\) we maintain a dynamic programming table
  \begin{equation}
    D_i \colon \big((X_i \to \{0, \ldots, \varrho\} \cup \{\infty\}) \times 2^{X_i \cap \vcust} \times \{0, \ldots, k\} \times \{0, \ldots, p\}\big) \to \{0, 1\}.
  \end{equation}
  We may refer to the value~\(0\) in the dynamic programming table as \emph{false} and to~\(1\) as \emph{true}.
  Let~\(i\) be a node of a nice tree decomposition, recall that~\(V_i\) is the set of vertices~\(u \in V(G)\) such that there exists a bag corresponding to a node in the subtree rooted at~\(i\) which contains~\(u\).
  For a distance labelling function~\(\dl\) of~\(X_i\) where~\(i \in V(T)\), a subset of clients~\(S \subseteq X_i \cap \vcust\) and constants~\(k_i\) and~\(p_i\) the entry~\(D_i[\dl, S, k_i, p_i]\) is going to be~\(1\) if and only if there exists a distance labelling function~\(\dl_*\) of~\(G[V_i]\) such that
  \begin{itemize}
    \item \(\dl_*\) \emph{agrees} with~\(\dl\) on~\(X_i\), i.e.~\((\forall u \in X_i)(\dl(u) = \dl_*(u))\),
    \item the cost of~\(\dl_*\) is~\(k_i\), that is \(\abs{\dl_*^{-1}(0)} = k_i\),
    \item the penalty of~\(\dl_*\) is~\(p_i\), that is \(\abs{\dl_*^{-1}(\infty) \cap \vcust} = p_i\),
    \item aside from clients~\(\dl_*^{-1}(\infty) \cap \vcust\) with an infinite label, every client~\(((V_i \setminus X_i) \cup S) \cap \vcust\) is satisfied.
  \end{itemize}
  In combination with Lemma~\ref{lem:kswo-epas-tw--dl}, the input admits a feasible solution of cost~\(\varrho\) if there exists a distance labelling~\(\dl_r\) of the root node~\(r\), and constants~\(k_r \leq k\) and~\(p_r \leq p\) such that the entry~\(D_r[\dl_r, X_r \cap \vcust, k_r, p_r]\) has value~\(1\).

  To simplify the exposition, assume that every value of the dynamic programming table at each node is initialized to value~\(0\).
  See Algorithm~\ref{alg:kswo-xp-alg-tw} for the pseudocode.

  \paragraph{Leaf node.}
  For a leaf node~\(\ell\) we have~\(V_\ell = \emptyset\).
  Thus the only table entry which can have value~\(1\) is~\(D_\ell[\dl, \emptyset, 0, 0]\) where the domain of~\(\dl\) is an empty set.

  \paragraph{Introduce node.}
  Let~\(i\) be an introduce node with a child node~\(j\) where~\(X_i = X_j \cup \{u\}\) and~\(u \not\in X_j\).
  Let~\(D_j[\dl_j, S_j, k_j, p_j]\) be an entry of the table of the child node~\(j\) with value~\(1\).
  We construct a distance labelling function~\(\dl_i\) which agrees with~\(\dl_j\) on~\(X_j\) and tries all possible values for~\(u\).
  In particular the constructed distance labelling functions set~\(\dl_i(u)\) to values~\(\{1, \ldots, \varrho\} \cup \{\infty\}\) and additionally we try the label~\(0\) for~\(u\) if~\(u\) is a supplier.
  For such a distance labelling function~\(\dl_i\) we compute~\(S_i\) to be the set of satisfied clients of~\(X_i\) as follows.
  We add the entire set~\(S_j\) to~\(S_i\), we add~\(u\) to~\(S_i\) if there exists a neighbor~\(v \in N(u)\) so that~\(\dl_i(u) \geq \dl_i(v) + d(u, v)\), and we add neighbors~\(w \in N(u)\) to~\(S_i\) for which it holds that~\(\dl_i(w) \geq \dl_i(u) + d(u, w)\).
  If~\(\dl_i(u) = 0\) and~\(k_j \leq k - 1\), then we set the entry~\(D_i[\dl_i, S_i, k_j + 1, p_j]\) to true.
  If~\(\dl_i(u) \in \{1, \ldots, \varrho\}\), then we set the entry~\(D_i[\dl_i, S_i, k_j, p_j]\) to true.
  If~\(\dl_i(u) = \infty\), \(u \in \vcust\), and~\(p_j \leq p - 1\), then we set the entry~\(D_i[\dl_i, S_i, k_j, p_j + 1]\) to true.
  If~\(\dl_i(u) = \infty\) and~\(u \in \vsup\), then we set the entry~\(D_i[\dl_i, S_i, k_j, p_j]\) to true.

  We proceed to showing the correctness.
  Let~\(\widehat{\dl}\) be a distance labelling function of~\(G[V_i]\) with cost~\(\widehat{k}\) and penalty~\(\widehat{p}\) such that all clients of~\(V_i \setminus X_i\) with a finite label are satisfied.
  We denote by~\(\widehat{S}\) the set of clients in~\(V_i\) satisfied by~\(\widehat{\dl}\).
  We want to show that the algorithm sets the table entry~\(D_i[\dl_i, S_i, k_i, p_i]\) to~\(1\) where~\(\dl_i\) agrees with~\(\widehat{\dl}\) on~\(X_i\), \(S_i = \widehat{S} \cap X_i\), \(k_i = \widehat{k}\), and~\(p_i = \widehat{p}\).
  By the induction hypothesis this property holds in the child node, thus there exists an entry~\(D_j[\dl_j, S_j, k_j, p_j]\) with value~\(1\) where~\(\dl_j\) agrees with~\(\widehat{\dl}\) on~\(X_j\), \(S_j = \widehat{S} \cap X_j\), and the cost and the penalty of~\(\widehat{\dl}\) restricted to~\(V_j\) is~\(k_j\) and~\(p_j\) respectively.
  The algorithm considers a distance labelling function~\(\dl_i\) which agrees with~\(\dl_j\) on~\(X_j\) and sets~\(\dl_i(u) = \widehat{\dl}(u)\).
  We need to show that the algorithm computes the set~\(S_i\) correctly.
  Recall that by Property~\ref{defi:tree-decomposition--prop-3} of tree decompositions we have~\(u \not\in V_j\).
  If a client of~\(X_i \setminus N[u]\) is satisfied by~\(\dl_j\), then it is also satisfied by~\(\dl_i\) since the label of the neighbor which satisfies it remains the same.
  Hence we have~\(S_j \subseteq S_i\) and~\(S_i \setminus S_j \subseteq N[u]\).
  The algorithm handles this by adding the entire set~\(S_j\) to~\(S_i\).
  If~\(u \in \widehat{S}\), then there exists a neighbor~\(v \in N(u) \cap V_i\) such that~\(\widehat{\dl}(u) \geq \widehat{\dl}(v) + d(u, v)\).
  In particular this means that there exists a node in a subtree rooted at~\(i\) whose bag contains~\(v\).
  We need to show that~\(v \in X_i\) as well.
  For a vertex~\(a \in V(G)\), let~\(T_a\) be the set of nodes of the tree decomposition~\(\mathcal{T}\) whose bags contain~\(a\).
  For a subset~\(U \subseteq V(T)\) we denote~\(X_U = \{X_w \mid w \in U\}\).
  By Property~\ref{defi:tree-decomposition--prop-2} of tree decompositions, there must exist a node of~\(\mathcal{T}\) whose corresponding bag contains both~\(u\) and~\(v\).
  Thus~\(X_{T_u} \cap X_{T_v} \neq \emptyset\).
  If it were the case that~\(v \not\in X_i\), then a tree decomposition satisfying properties~\(u \in X_i\), \(u \not\in X_j\), \(v \in V_i\), and that there exists a node whose bag contains both~\(u\) and~\(v\) necessarily violates Property~\ref{defi:tree-decomposition--prop-3}.
  We conclude that~\(v \in X_i\).
  Since~\(\widehat{\dl}\) agrees with~\(\dl_i\), the algorithm will find~\(v\) in~\(X_i\) and add~\(u\) to~\(S_i\).
  For a client~\(w \in N(u) \cap X_i\), we have~\((N(w) \cap X_i) \setminus (N(w) \cap X_j) = \{u\}\), since~\(i\) is an introduce node introducing~\(u\).
  Therefore the only reason why~\(w\) would be satisfied by~\(\dl_i\) but not by~\(\dl_j\) is that it is satisfied by~\(u\).
  Then the set~\(S_j\) is the set~\(S_i\) without clients in~\(X_i\) which are satisfied only by~\(u\).
  The algorithm adds to~\(S_i\) those vertices of~\(N(u) \cap X_i\) which are satisfied by~\(u\).
  It remains to describe values~\(k_i\) and~\(p_i\), we distinguish the following cases:
  \begin{itemize}
    \item \textbf{Case~\(\widehat{\dl}(u) = 0\).}
      As~\(u \not\in V_j\), we have~\(\abs{\widehat{\dl}^{-1}(0) \cap V_i} = \abs{\widehat{\dl}^{-1}(0) \cap V_j} + 1\).
      Since~\(\widehat{\dl}(u) \neq \infty\), we have~\(\abs{\widehat{\dl}^{-1}(\infty) \cap V_i \cap \vcust} = \abs{\widehat{\dl}^{-1}(\infty) \cap V_j \cap \vcust}\).
      Thus the algorithm sets~\(k_i = k_j + 1\) and~\(p_i = p_j\).
    \item \textbf{Case~\(\widehat{\dl}(u) \in \{1, \ldots, \varrho\}\).}
      Since~\(\widehat{\dl}(u) \not\in \{0, \infty\}\), we have~\(\abs{\widehat{\dl}^{-1}(0) \cap V_i} = \abs{\widehat{\dl}^{-1}(0) \cap V_j}\) and~\(\abs{\widehat{\dl}^{-1}(\infty) \cap V_i \cap \vcust} = \abs{\widehat{\dl}^{-1}(\infty) \cap V_j \cap \vcust}\), hence~\(k_i = k_j\) and~\(p_i = p_j\) respectively.
    \item \textbf{Case~\(\widehat{\dl}(u) = \infty\).}
      We have~\(\abs{\widehat{\dl}^{-1}(0) \cap V_i} = \abs{\widehat{\dl}^{-1}(0) \cap V_j}\) and therefore~\(k_i = k_j\).
      If~\(u \in \vcust\), then~\(\abs{\widehat{\dl}^{-1}(\infty) \cap V_i \cap \vcust} = \abs{\widehat{\dl}^{-1}(\infty) \cap V_j \cap \vcust} + 1\) and~\(p_i = p_j + 1\).
      Otherwise~\(u \in \vsup\) and then we have~\(p_i = p_j\).
  \end{itemize}

  For the opposite direction, assume that the algorithm sets the value of an entry~\(D_i[\dl_i, S_i, k_i, p_i]\) to~\(1\).
  By the description of the algorithm, this means that there exists an entry~\(D_j[\dl_j, S_j, k_j, p_j]\) in the table of the child node set to~\(1\) where~\(\dl_i\) agrees with~\(\dl_j\) on~\(X_j\), \(S_j \subseteq S_i\), \(k_j \leq k_i\), and~\(p_j \leq p_i\).
  By the induction hypothesis there exists a valid distance labelling function~\(\widehat{\dl}_j\) on~\(V_j\) which agrees with~\(\dl_j\) on~\(X_j\), satisfies clients~\(S_j\) of~\(X_j\), and has cost~\(k_j\) and penalty~\(p_j\).
  We claim that the function~\(\widehat{\dl}_i\) which extends~\(\widehat{\dl}_j\) by setting~\(\widehat{\dl}_i(u) = \dl_i(u)\) has the required properties.
  We need to show that all clients of~\(S_i\) are satisfied by~\(\widehat{\dl}_i\).
  Every satisfied client of~\(S_j\) is still satisfied by~\(\widehat{\dl}_i\) since the distance label of the neighbor which satisfies it is preserved.
  From the description of the algorithm it is clear that the added clients~\(S_i \setminus S_j\) are satisfied by~\(\widehat{\dl}_i\).
  It remains to verify that the algorithm computes the cost of~\(\widehat{\dl}_i\) correctly.
  Recall that~\(V_i \setminus V_j = \{u\}\).
  We distinguish the following cases based on the description of the algorithm:
  \begin{itemize}
    \item \textbf{Case~\(\dl_i(u) = 0\).}
      In this case~\(\widehat{\dl}_i^{-1}(0) = \widehat{\dl}_j^{-1}(0) \cup \{u\}\) and~\(\widehat{\dl}_i^{-1}(\infty) \cap \vcust = \widehat{\dl}_j^{-1}(\infty) \cap \vcust\).
      Then the cost of~\(\widehat{\dl}_i\) is~\(k_j + 1\) and the penalty is~\(p_j\).
    \item \textbf{Cases~\(\dl_i(u) \in \{1, \ldots, \varrho\}\) and~\(\dl_i(u) = \infty \wedge u \in \vsup\).}
      In this case~\(\widehat{\dl}_i^{-1}(0) = \widehat{\dl}_j^{-1}(0)\) and~\(\widehat{\dl}_i^{-1}(\infty) \cap \vcust = \widehat{\dl}_j^{-1}(\infty) \cap \vcust\).
      Then the cost of~\(\widehat{\dl}_i\) is~\(k_j\) and the penalty is~\(p_j\).
    \item \textbf{Case~\(\dl_i(u) = \infty \wedge u \in \vcust\).}
      In this case~\(\widehat{\dl}_i^{-1}(0) = \widehat{\dl}_j^{-1}(0)\) and~\(\widehat{\dl}_i^{-1}(\infty) \cap \vcust = \left(\widehat{\dl}_j^{-1}(\infty) \cap \vcust\right) \cup \{u\}\).
      Then the cost of~\(\widehat{\dl}_i\) is~\(k_j\) and the penalty is~\(p_j + 1\).
  \end{itemize}
  
  \paragraph{Forget node.}
  Let~\(i\) be a forget node where~\(X_i = X_j \setminus \{u\}\) and~\(u \in X_j\).
  For any distance labelling function~\(\dl\) of~\(X_i\) we denote~\(L_c(\dl)\) the set of functions which extend~\(\dl\) by assigning labels~\(\{1, \ldots, \varrho\}\) to~\(u\), \(L_s(\dl)\) the set of functions which extend~\(\dl\) by assigning labels~\(\{0, \ldots, \varrho\} \cup \{\infty\}\) to~\(u\), and \(L_\infty(\dl)\) the extension of~\(\dl\) which assigns label~\(\infty\) to~\(u\).
  Let~\(\dl_i\) be any distance labelling function of~\(X_i\), \(S_i\)~be any subset of clients of~\(X_i\), \(k_i \in \{0, \ldots, k\}\), and~\(p_i \in \{0, \ldots, p\}\).
  If~\(u\) is a supplier, then we set
  \begin{equation}
    D_i[\dl_i, S_i, k_i, p_i] = \bigvee_{\dl_j \in L_s(\dl_i)} D_j[\dl_j, S_i, k_i, p_i].
  \end{equation}
  If~\(u\) is a client, then we set
  \begin{equation}
    D_i[\dl_i, S_i, k_i, p_i] = \left(\bigvee_{\dl_j \in L_c(\dl_i)} D_j[\dl_j, S_i \cup \{u\}, k_i, p_i]\right) \vee D_j[L_\infty(\dl), S_i, k_i, p_i].
  \end{equation}

  We proceed to showing the correctness.
  Recall that by properties of nice tree decompositions, we have~\(V_i = V_j\) for the forget node~\(i\).
  Let~\(\widehat{\dl}\) be a distance labelling function of~\(G[V_i]\) of cost~\(\widehat{k}\) and penalty~\(\widehat{p}\) such that all clients of~\(V_i \setminus X_i\) with a finite label are satisfied.
  Given a node~\(t\) of the tree decomposition, let~\(\widehat{\dl}_t\) be the restriction of~\(\widehat{\dl}\) to~\(X_t\), and~\(\widehat{S}_t\) be the set of satisfied of clients of~\(X_t\) by~\(\widehat{\dl}\).
  Our goal is to show that the algorithm sets the entry~\(D_i[\widehat{\dl}_i, \widehat{S}_i, \widehat{k}, \widehat{p}]\) to~\(1\).
  From the induction hypothesis, the entry~\(D_j[\widehat{\dl}_j, \widehat{S}_j, k_j, p_j]\) is set to~\(1\) for some values~\(k_j\) and~\(p_j\); in fact it is the case that~\(k_j = \widehat{k}\) and~\(p_j = \widehat{p}\) as~\(V_i = V_j\).
  Since the algorithm inspects all entries of the table~\(D_i\), it will eventually consider the entry~\(D_i[\widehat{\dl}_i, \widehat{S}_i, \widehat{k}, \widehat{p}]\).
  By trying all possible values for the forgotten vertex~\(u\), it will consider the labelling~\(\widehat{\dl}_j\) of~\(X_j\).
  To finish the proof of this direction, it remains to determine the relationship between~\(\widehat{S}_i\) and~\(\widehat{S}_j\).
  \begin{itemize}
    \item \textbf{Case~\(u \in \vsup\).}
      We have~\(X_i \cap \vcust = X_j \cap \vcust\) and thus~\(\widehat{S}_i = \widehat{S}_j\).
    \item \textbf{Case~\(\widehat{\dl}(u) \in \{1, \ldots, \varrho\}\) and~\(u \in \vcust\).}
      If the label of the forgotten client~\(u\) is finite, then it must be satisfied by~\(\widehat{\dl}\) from the requirement that~\(\widehat{\dl}\) satisfies all vertices of~\(V_i \setminus X_i\).
      Thus~\(\widehat{S}_i \cup \{u\} = \widehat{S}_j\).
    \item \textbf{Case~\(\widehat{\dl}(u) = \infty\) and~\(u \in \vcust\).}
      If the label of the forgotten client~\(u\) is~\(\infty\), then it is ignored by~\(\widehat{\dl}\).
      In this case we have~\(\widehat{S}_i = \widehat{S}_j\).
  \end{itemize}

  For the opposite direction let~\(D_i[\dl_i, S_i, k_i, p_i]\) be an entry set to~\(1\) by the algorithm.
  Then it follows from the description of the algorithm, that there is an entry~\(D_j[\dl_j, S_j, k_j, p_j]\) set to~\(1\) in the table of the child~\(j\) where~\(\dl_j\) is an extension of~\(\dl_i\) to~\(X_j\), \(S_j \supseteq S_i\), \(k_i = k_j\) and~\(p_i = p_j\).
  By the induction hypothesis, there exists a distance labelling function~\(\widehat{\dl}_j\) on~\(V_j\) which agrees with~\(\dl_j\) on~\(X_j\), satisfies clients~\(S_j\) of~\(X_j\), has cost~\(k_j\) and penalty~\(p_j\).
  We claim that~\(\widehat{\dl}_j\) also agrees with~\(\dl_i\), satisfies clients~\(S_i\) of~\(X_i\), has cost~\(k_i\) and penalty~\(p_i\).
  Since~\(\dl_j\) agrees with~\(\dl_i\) on~\(X_i\) and~\(X_i \subseteq X_j\), \(\widehat{\dl}_j\) agrees with~\(\dl_i\).
  The property that every client of~\(S_i\) is satisfied by~\(\widehat{\dl}_j\) follows from~\(S_i \subseteq S_j\), \(V_i = V_j\), and the fact that~\(\widehat{\dl}_j\) satisfies~\(S_j\) by the induction hypothesis.
  From~\(V_i = V_j\) it also follows that~\(k_i = k_j\) and~\(p_i = p_j\).

  \paragraph{Join node.}
  Let~\(i\) be a join node with children~\(j_1\) and~\(j_2\) where~\(X_i = X_{j_1} = X_{j_2}\).
  Let~\(\dl_i\) be a distance labelling function on~\(X_i\), then for each pair of true entries~\(D_{j_1}[\dl, S_1, k_1, p_1]\) and~\(D_{j_2}[\dl, S_2, k_2, p_2]\) we set to~\(1\) the entry
  \begin{equation}
    D_i\left[\dl_i, S_1 \cup S_2, k_1 + k_2 - \abs{\dl_i^{-1}(0)}, p_1 + p_2 - \abs{\dl_i^{-1}(\infty) \cap \vcust}\right].
  \end{equation}

  We proceed to showing the correctness.
  It follows from the properties of nice tree decompositions that~\(V_i = V_{j_1} \cup V_{j_2}\).
  Let~\(\widehat{\dl}\) be a valid distance labelling function of~\(G[V_i]\) of cost~\(\widehat{k}\) and penalty~\(\widehat{p}\) such that all clients of~\(V_i \setminus X_i\) with a finite label are satisfied.
  We denote by~\(\widehat{S}_i\) the set of satisfied clients in~\(X_i\) and by~\(\widehat{\dl}_i\) the restriction of~\(\widehat{\dl}\) to~\(X_i\).
  Let~\(a \in \{1, 2\}\), \(\widehat{\dl}_a\) be the restriction of~\(\widehat{\dl}\) to~\(V_{j_a}\), \(\widehat{S}_{j_a}\)~be the set of clients of~\(X_{j_a}\) satisfied by~\(\widehat{\dl}_a\), \(\widehat{C}_a\)~be the set of opened suppliers in~\(V_{j_a}\), and~\(\widehat{O}_a\) be the set of outliers in~\(V_{j_a}\).
  Note that~\(\widehat{\dl}, \widehat{\dl}_1\) and~\(\widehat{\dl}_2\) agree with each other on~\(X_i\).
  From the induction hypothesis the entries~\(D_{j_a}\left[\widehat{\dl}_i, \widehat{S}_{j_a}, \abs{\widehat{C}_{j_a}}, \abs{\widehat{O}_{j_a}}\right]\) are set to~\(1\).
  To show that~\(\widehat{S}_i = \widehat{S}_{j_1} \cup \widehat{S}_{j_2}\), we use the standard approach of showing~\(\widehat{S}_i \supseteq \widehat{S}_{j_1} \cup \widehat{S}_{j_2}\) and~\(\widehat{S}_i \subseteq \widehat{S}_{j_1} \cup \widehat{S}_{j_2}\).
  From~\(V_{j_1} \subseteq V_i\) it trivially follows that~\(\widehat{S}_{j_a} \subseteq \widehat{S}_i\).
  For a satisfied client~\(v \in \widehat{S}_i\), we want to show that~\(v \in \widehat{S}_{j_1} \cup \widehat{S}_{j_2}\).
  As~\(V_i = V_{j_1} \cup V_{j_2}\), the neighbor~\(w\) which satisfies~\(v\) lies in~\(V_{j_1} \cup V_{j_2}\).
  Hence it must be the case that~\(v\) is satisfied in at least in one of~\(V_{j_1}\) or~\(V_{j_2}\) and thus~\(v\) is satisfied in at least one of~\(\widehat{S}_{j_1}\) or~\(\widehat{S}_{j_2}\).
  It follows from Property~\ref{defi:tree-decomposition--prop-3} of tree decompositions that if a vertex~\(w\) is simultaneously contained in a bag of some node of the subtree rooted in~\(j_1\) and in a bag of some node of the subtree rooted in~\(j_2\), then~\(w \in X_i\).
  This means that~\(V_{j_1} \cap V_{j_2} \subseteq X_i\) and in particular~\(\widehat{C}_1 \cap \widehat{C}_2 \subseteq X_i\) and~\(\widehat{O}_1 \cap \widehat{O}_2 \subseteq X_i\).
  Since~\(\widehat{\dl}\), \(\widehat{\dl}_1\), and~\(\widehat{\dl}_2\) agree with each other on~\(X_i\), \(\widehat{C}_1 \cap X_i = \widehat{C}_2 \cap X_i\) and~\(\widehat{O}_1 \cap X_i = \widehat{O}_2 \cap X_i\).
  Using these facts we have~\(\abs{\widehat{C}_1} + \abs{\widehat{C}_2} = \widehat{k} + \abs{\widehat{\dl}^{-1}(0) \cap X_i}\) and~\(\abs{\widehat{O}_1} + \abs{\widehat{O}_2} = \widehat{p} + \abs{\widehat{\dl}^{-1}(\infty) \cap X_i \cap \vcust}\), as every opened supplier and every outlier is counted twice in~\(X_i\).

  For the opposite direction, let~\(D_i[\dl_i, S_i, k_i, p_i]\) be an entry set to~\(1\) by the algorithm.
  From the description of the algorithm, this means that there exist entries~\(D_{j_1}[\dl_i, S_{j_1}, k_{j_1}, p_{j_1}]\) and~\(D_{j_2}[\dl_i, S_{j_2}, k_{j_2}, p_{j_2}]\) set to~\(1\) where~\(S_{j_1}\) and~\(S_{j_2}\) are subsets of~\(X_i\) (where~\(X_i = X_{j_1} = X_{j_2}\) for a join node~\(i\)).
  By the induction hypothesis, for~\(a \in \{1, 2\}\) there exists a valid distance labelling function~\(\widehat{\dl}_a\) on~\(V_{j_a}\) which agrees with~\(\dl_i\) on~\(X_{j_a}\), satisfies clients~\(S_{j_a} \subseteq X_{j_a}\), and has cost~\(\widehat{k}_a\) and penalty~\(\widehat{p}_a\).
  We claim that a function~\(\dl\) which behaves as~\(\widehat{\dl}_1\) on~\(V_{j_1}\) and as~\(\widehat{\dl}_2\) on~\(V_{j_2}\) has the desired properties.
  To verify that~\(\dl\) is a function (and not a multifunction), we use the fact that~\(V_{j_1} \cap V_{j_2} \subseteq X_i\).
  The algorithm requires that~\(\widehat{\dl}_1\) and~\(\widehat{\dl}_2\) behave the same on~\(X_i\), thus~\(\dl\) is indeed a function.
  We can prove the remaining required properties, i.e. \(S_i = S_{j_1} \cup S_{j_2}\), \(k_i = k_{j_1} + k_{j_2} - \abs{\dl_i^{-1}(0)}\), and~\(p_i = p_{j_1} + p_{j_2}\), identically as in the proof of the opposite direction.

  \paragraph{Running time.}
  The size of the table at each node is at most~\((\varrho + 2)^\tw \cdot 2^\tw \cdot (k + 1) \cdot (p + 1)\).
  In each introduce node~\(i\), we inspect all table entries of its child~\(j\) and for each such child entry we recalculate the set of satisfied vertices, suppliers with label~\(0\), and ignored clients in time~\(\tw\).
  In each forget node~\(i\), for each table entry of~\(i\) we inspect up to~\(\varrho + 2\) entries in the table of the child~\(j\).
  In each join node~\(i\), we try to combine all possible entries from its children~\(j_1\) and~\(j_2\) which takes time~\(\abs{D_{j_1}} \cdot \abs{D_{j_2}} \leq ((\varrho + 2)^\tw \cdot 2^\tw \cdot (k + 1) \cdot (p + 1))^2\).
  Overall, the running time of the algorithm is at most~\(\bigO^*((4(\varrho + 2))^{\bigO(\tw)})\).
  \qed
\end{proof}

\begin{algorithmic}[1]
  \Statex \textbf{XP algorithm for \problemName{kSwO}: Dynamic programming on tree decompositions}
  \Function{\textproc{LeafNode}}{\ensuremath{\ell}}
    \State Let~\(\dl\) be the distance labelling function with an empty domain
    \State \(D_\ell[\dl, \emptyset, 0, 0] \gets 1\)
  \EndFunction

  \Statex \vspace{0.5ex} 
  \Function{\textproc{IntroduceNode}}{\ensuremath{i}}
    \State Let \(j\) be the child of \(i\) with \(X_i = X_j \cup \{u\}\) and \(u\notin X_j\)
    \ForAll{entries \((\dl_j,S_j,k_j,p_j)\) with \(D_j[\dl_j,S_j,k_j,p_j]=1\)}
      \State \(\mathrm{Vals}\gets \langle\varrho\rangle\cup\{\infty\}\cup\bigl(\{0\}\ \text{if } u\in\vsup \text{ else }\varnothing\bigr)\)
      \State Initialize~\(\dl_i\) to be the same as~\(\dl_j\) on~\(X_j\)
      \ForAll{\(a\in\mathrm{Vals}\)}
        \State \(\dl_i(u) \gets a\)
        \State \(S_i \gets S_j\)
        \If{\(\exists v\in N(u) \cap X_i:\ \dl_i(u)\ge \dl_i(v)+d(u,v)\)}
          \State \(S_i \gets S_i\cup\{u\}\)
        \EndIf
        \ForAll{\(w\in N(u)\cap X_i\)}
          \If{\(\dl_i(w)\ge \dl_i(u)+d(u,w)\)}
            \State \(S_i \gets S_i\cup\{w\}\)
          \EndIf
        \EndFor
        \If{\(a=0\) \textbf{and} \(k_j\le k-1\)}
          \State \(D_i[\dl_i,S_i,\,k_j{+}1,\,p_j]\gets 1\)
        \ElsIf{\(a\in\langle\varrho\rangle\)}
          \State \(D_i[\dl_i,S_i,\,k_j,\,p_j]\gets 1\)
        \ElsIf{\(a=\infty\) \textbf{and} \(u\in\vcust\) \textbf{and} \(p_j\le p-1\)}
          \State \(D_i[\dl_i,S_i,\,k_j,\,p_j{+}1]\gets 1\)
        \ElsIf{\(a=\infty\) \textbf{and} \(u\in\vsup\)}
          \State \(D_i[\dl_i,S_i,\,k_j,\,p_j]\gets 1\)
        \EndIf
      \EndFor
    \EndFor
  \EndFunction

  \Statex \vspace{0.5ex} 
  \Function{ForgetNode}{\ensuremath{i}}
    \State Let \(j\) be the child of \(i\) with \(X_i = X_j \setminus \{u\}\) and \(u \in X_j\)
    \State Given a distance labelling function~\(\dl\) of~\(X_i\), \(L_c(\dl)\) is the set of functions which extend~\(\dl\) by assigning labels~\(\langle\varrho\rangle\) to~\(u\)
    \State Given a distance labelling function~\(\dl\) of~\(X_i\), \(L_s(\dl)\) is the set of functions which extend~\(\dl\) by assigning labels~\(\langle 0, \varrho\rangle \cup \{\infty\}\) to~\(u\)
    \State Given a distance labelling function~\(\dl\) of~\(X_i\), \(L_\infty(\dl)\) is the extension of~\(\dl\) which assigns~\(\infty\) to~\(u\)
    \ForAll{distance labelling functions~\(\dl_i\) on \(X_i\)}
      \ForAll{\(S_i \subseteq X_i \cap \vcust\)}
        \ForAll{\(k_i \in \langle 0, k\rangle\), \(p_i \in \langle 0, p\rangle\)}
          \If{\(u \in \vsup\)}

            \State \(D_i[\dl_i, S_i, k_i, p_i] \gets \bigvee_{\dl_j \in L_s(\dl_i)} D_j[\dl_j, S_i, k_i, p_i].\)
          \Else
            \State \(D_i[\dl_i, S_i, k_i, p_i] \gets \left(\bigvee_{\dl_j \in L_c(\dl_i)} D_j[\dl_j, S_i \cup \{u\}, k_i, p_i]\right) \vee D_j[L_\infty(\dl), S_i, k_i, p_i]\)
          \EndIf
        \EndFor
      \EndFor
    \EndFor
  \EndFunction

  \Statex \vspace{0.5ex}
  \Function{JoinNode}{\ensuremath{i}}
    \State Let \(j_1,j_2\) be the children of \(i\) with \(X_i = X_{j_1} = X_{j_2}\)
    \ForAll{distance labelling functions~\(\dl_i\) on \(X_i\)}
      \State \(Z \gets \{v \in X_i : \dl_i(v) = 0\}\)
      \State \(I \gets \{v \in X_i \cap \vcust : \dl_i(v) = \infty\}\)
      \ForAll{entries \((\dl, S_1, k_1, p_1)\) with \(D_{j_1}[\dl, S_1, k_1, p_1] = 1\) and \(\dl = \dl_i\)} \Comment{match labels}
        \ForAll{entries \((\dl', S_2, k_2, p_2)\) with \(D_{j_2}[\dl', S_2, k_2, p_2] = 1\) and \(\dl' = \dl_i\)}
          \State \(S_i \gets S_1 \cup S_2\)
          \State \(k_i \gets k_1 + k_2 - \abs{Z}\)
          \State \(p_i \gets p_1 + p_2 - \abs{I}\)
          \If{\(k_i \le k \wedge p_i \le p\)}
            \State \(D_i[\dl_i, S_i, k_i, p_i] \gets 1\)
          \EndIf
        \EndFor
      \EndFor
    \EndFor
  \EndFunction
\end{algorithmic}
\captionof{algorithm}{Dynamic programming on tree decompositions for an XP algorithm for~\problemName{kSwO} (Theorem~\ref{thm:kswo-xp-alg-tw})}
\label{alg:kswo-xp-alg-tw}

\subsubsection{The approximation scheme.}

Now we describe a parameterized approximation scheme based on the algorithm from Theorem~\ref{thm:kswo-xp-alg-tw}.
We will need the following result by Chatterjee et al.~\cite{chatterjee2014optimal}.
\begin{theorem}[{\cite{chatterjee2014optimal}}]
  \label{thm:tree-decomposition-balancing}
  Let~\(G\) be a graph.
  There exists an algorithm which, given a tree decomposition~\(\mathcal{T}\) of~\(G\) such that~\(\mathcal{T}\) has~\(n\) nodes and width~\(\tw\), produces a nice tree decomposition of~\(G\) with width at most~\(4\tw + 3\) and height~\(\bigO(\tw \cdot \log n)\) in time~\(\bigO(\tw \cdot n)\).
\end{theorem}

Let us give an approximate version of the distance labelling problem for a fixed error parameter~\(\varepsilon > 0\).
This is a generalization of the approximate distance labelling used in the original algorithm~\cite{KATSIKARELIS201990}.
Let~\((G, k, p)\) be a \problemName{kSwO} instance with edge lengths~\(d: E \to \N\) and~\(\delta > 0\) some appropriately chosen secondary parameter (we will eventually set~\(\delta \approx \frac{\varepsilon}{\log n}\)).
Let~\(\Sigma_\varrho = \{(1 + \delta)^i: i \in \N, (1 + \delta)^i \leq (1 + \varepsilon)\varrho\}\), we define a~\emph{\(\delta\)\nobreakdash-labelling function} of~\(V\) as a function~\(\dl_\delta: V \to \Sigma_\varrho \cup \{0, \infty\}\).
We require that only suppliers can have label~\(0\).
A vertex~\(u\) is~\emph{\(\varepsilon\)\nobreakdash-satisfied} if~\(\dl_\delta(u) = 0\) or if~\(u\) has a finite label and there exists~\(v \in N(u)\) such that~\(\dl_\delta(u) \geq \dl_\delta(v) + \frac{d(u, v)}{1 + \varepsilon}\).
Given a~\(\delta\)\nobreakdash-labelling function~\(\dl_\delta\), if every client is either \(\varepsilon\)\nobreakdash-satisfied or has label~\(\infty\), then we say that~\(\dl_\delta\) is \emph{valid}.
We define the~\emph{cost} of such a function~\(\dl_\delta\) as~\(\abs{\dl_\delta^{-1}(0)}\) and its \emph{penalty} as~\(\abs{\dl_\delta^{-1}(\infty) \cap \vcust}\).
The following lemma shows that given a~\(\delta\)\nobreakdash-labelling function of cost~\(k\) and penalty~\(p\), we can produce a solution to the~\problemName{kSwO} problem which opens~\(k\)~centers, creates~\(p\)~outliers and has cost~\((1 + \varepsilon)^2\varrho\).
\begin{lemma}
  \label{lem:delta-labelling-equivalence}
  Let~\(\mathcal{I} = (G, k, p)\) be an instance of the \problemName{kSwO} problem.
  If there exists a valid~\(\delta\)\nobreakdash-labelling function of~\(G\) with cost at most~\(k\) and penalty at most~\(p\), then~\(\mathcal{I}\) has a feasible solution of cost~\((1 + \varepsilon)^2\varrho\).
\end{lemma}
\begin{proof}
  We select vertices with label~\(0\) as the solution~\(S\).
  The number of opened suppliers is the cost of the function, hence we open at most~\(k\)~suppliers.
  We will show by induction on~\(i\) that if~\(\dl_\delta(u) = (1 + \delta)^i\), then~\(\dist(u, S) \leq (1 + \varepsilon)\dl_\delta(u)\).
  In the base case let~\(u\) be an \(\varepsilon\)\nobreakdash-satisfied vertex with~\(\dl_\delta(u) = (1 + \delta)\).
  Then there exists a neighbor~\(v \in N(u)\) with~\(\dl_\delta(u) \geq \dl_\delta(v) + \frac{d(u, v)}{1 + \varepsilon}\).
  Since~\(d(u, v) > 0\), it follows that~\(\dl_\delta(u) > \dl_\delta(v)\) and the only possible \(\delta\)\nobreakdash-label less than~\((1 + \delta)\) is~\(0\), hence~\(\dl_\delta(v) = 0\).
  Then we have~\((1 + \delta) \geq \frac{d(u, v)}{1 + \varepsilon}\) which shows the base case as~\(v \in S\).
  For the induction step, let~\(u\) be an \(\varepsilon\)\nobreakdash-satisfied vertex with~\(\dl_\delta(u) = (1 + \delta)^{i + 1}\).
  There exists a neighbor~\(v \in N(u)\) such that~\(\dl_\delta(u) \geq \dl_\delta(v) + \frac{d(u, v)}{1 + \varepsilon}\).
  As edge lengths are positive, we have~\(\dl_\delta(u) > \dl_\delta(v)\) and using induction hypothesis we receive~\(\dist(v, S) \leq \dl_\delta(v)\).
  By triangle inequality we have~\(\dist(u, S) \leq d(u, v) + \dist(v, S) \leq d(u, v) + \dl_\delta(v) = (1 + \varepsilon)\left(\frac{d(u, v)}{1 + \varepsilon} + \dl_\delta(v)\right) \leq (1 + \varepsilon)\dl_\delta(u)\).

  Since all satisfied clients have a \(\delta\)\nobreakdash-label at most~\((1 + \varepsilon)\varrho\), all satisfied clients are at distance at most~\((1 + \varepsilon)^2\varrho\) from~\(S\).
  The number of outliers is at most~\(p\) as the penalty of~\(\dl_\delta\) is at most~\(p\).
  \qed
\end{proof}

The following two lemmas from~\cite{KATSIKARELIS201990}, which they use for their EPAS for \problemName{\(k\)\nobreakdash-Center}, will be useful for us as well.
The first lemma shows that adding all missing edges between vertices of a single bag of length equal to their shortest\nobreakdash-path distance does not change the set of solutions.
The original formulation is for \problemName{\(k\)\nobreakdash-Center}, however, their proof generalizes to \problemName{kSwO} without any major modifications.
\begin{lemma}[{\cite[Lemma~29]{KATSIKARELIS201990}}]
  \label{lem:tree-decomposition-metric-closure-in-bag}
  Let~\(\mathcal{I} = (G, k, p)\) be a \problemName{kSwO} instance, \(\mathcal{T}\) a tree decomposition of~\(G\) and~\(u, v \in V\) two vertices which appear together in a bag of~\(\mathcal{T}\) and~\((u, v) \not\in E\).
  Let~\(G'\) be the graph obtained from~\(G\) by adding the edge~\(\{u, v\}\) with length~\(\dist(u, v)\) and let~\(\mathcal{I}' = (G', k, p)\) be a \problemName{kSwO} instance.
  Then~\(\mathcal{I}\) has a feasible solution of cost~\(\varrho\) if and only if~\(\mathcal{I}'\) does.
\end{lemma}

The second lemma shows that an algorithm with a running time in the form~\(\bigO^*\left((\frac{\log n}{\varepsilon})^{\bigO(k)}\right)\) is still an \(\classFPT\) algorithm.
\begin{lemma}[{\cite[Lemma~1]{KATSIKARELIS201990}}]
  \label{lem:exp-in-log-n-is-fpt}
  Let~\(\mathcal{A}\) be an algorithm for a parameterized problem with parameter~\(k\) such that the running time of~\(\mathcal{A}\) is~\(\bigO^*\left((\frac{\log n}{\varepsilon})^{\bigO(k)}\right)\).
  Then the running time of~\(\mathcal{A}\) can be bounded by~\(\bigO^*\left((\frac{k}{\varepsilon})^{\bigO(k)}\right)\).
\end{lemma}

We are ready to prove Theorem~\ref{thm:kswo-epas-tw}.
The proof follows the proof of the original algorithm, cf.~\cite[Theorem~31]{KATSIKARELIS201990}.
The main obstacle lies in bounding the accumulated error during the execution of the algorithm.
The original proof heavily relies on the properties of the algorithm they present for \problemName{\(k\)\nobreakdash-Center}.
Hence, we will have to modify their proof to work with the algorithm we give for \problemName{kSwO}.

\begin{proof}[{Theorem~\ref{thm:kswo-epas-tw}}]
  Our algorithm will follow along the same lines as the algorithm in Theorem~\ref{thm:kswo-xp-alg-tw}.
  The major difference is that instead of distance labelling functions we consider~\(\delta\)\nobreakdash-labelling functions for some~\(\delta\) we specify later and instead of satisfiability we use~\(\varepsilon\)\nobreakdash-satisfiability.

  Recall that~\(\Sigma_\varrho = \{0\} \cup \{(1 + \delta)^i: i \in \N, (1 + \delta)^i \leq (1 + \varepsilon)\varrho\} \cup \{\infty\}\).
  For each node~\(i \in V(T)\) we define a table
  \begin{equation}
    D^\delta_i \colon \left((X_i \to \Sigma_\varrho) \times 2^{X_i \cap \vcust} \times \{0, \ldots, k\} \times \{0, \ldots, p\} \right) \to \{0, 1\}.
  \end{equation}
  We may refer to value~\(1\) in the dynamic programming table as~\emph{true} and to value~\(0\) as \emph{false}.
  Recall that our definitions of a~\(\delta\)\nobreakdash-labelling allows only suppliers to have label~\(0\).
  For a node~\(i \in V(T)\), a~\(\delta\)\nobreakdash-labelling function~\(\dl_\delta \colon X_i \to \Sigma_\varrho\), a subset of clients~\(S \subseteq 2^{X_i \cap \vcust}\), and integers~\(k_i\) and~\(p_i\) where~\(0 \leq k_i \leq k\) and~\(0 \leq p_i \leq p\), the value of an entry~\(D^\delta_i[\dl_\delta, S, k_i, p_i]\) is~\(1\) if and only if there exists a~\(\delta\)\nobreakdash-labelling of~\(G[V_i]\) which agrees with~\(\dl_\delta\) on~\(X_i\), satisfies clients~\(((V_i \setminus X_i) \cup S) \cap \vcust\), has cost~\(k_i\) and penalty~\(p_i\).

  For the rest of the proof, we denote by~\(n\) the number of nodes of the tree decomposition~\(\mathcal{T}\) provided on input.
  We start by preprocessing the graph using Theorem~\ref{thm:tree-decomposition-balancing} and Lemma~\ref{lem:tree-decomposition-metric-closure-in-bag}.
  We obtain a tree decomposition~\(\mathcal{T}'\) of the input graph of width~\(4\tw + 3\) and height~\(H\) where~\(H \in \bigO(\tw \cdot \log n)\).
  For every pair of vertices~\(u, v\) which appear together in some bag of~\(\mathcal{T}'\) we have an edge with length at most their shortest\nobreakdash-path distance.
  We define the~\emph{height} of a node of~\(\mathcal{T}'\) inductively where the height of a leaf node is~\(1\) and the height of any inner node is~\(1\) plus the maximum of the heights of its children.
  Under this definition the root node has height~\(H\) and all other bags have height less than~\(H\).
  We may refer to a height of a bag by which we mean the height of the node corresponding to the bag.

  We set~\(\delta = \frac{\varepsilon}{2H} = \Omega\big(\frac{\varepsilon}{\tw \cdot \log n}\big)\).
  Observe that this choice of~\(\delta\) gives for all~\(h \leq H\) that~\((1 + \delta)^h \leq \left(1 + \frac{\varepsilon}{2H}\right)^H \leq e^{\varepsilon/2} \leq 1 + \varepsilon\) for sufficiently small~\(\varepsilon\) (it suffices to assume without loss of generality that~\(\varepsilon < \frac{1}{4}\)).
  The goal is to return a feasible solution of cost~\((1 + \varepsilon)^2\varrho\) if a feasible solution of cost~\(\varrho\) exists by producing a~\(\delta\)\nobreakdash-labelling and invoking Lemma~\ref{lem:delta-labelling-equivalence}.
  The approximation ratio can then be reduced to~\(1 + \varepsilon\) by adjusting~\(\varepsilon\) appropriately.

  We now present the dynamic programming procedure.
  It only differs from the algorithm in Theorem~\ref{thm:kswo-xp-alg-tw} by considering~\(\delta\)\nobreakdash-labelling functions instead of distance labelling functions and~\(\varepsilon\)\nobreakdash-satisfiability instead of satisfiability.

  \begin{itemize}
    \item \textbf{Leaf node.}
      For a leaf node~\(\ell\) we have~\(V_\ell = \emptyset\).
      Thus the only true entry is~\(D^\delta_\ell[\dl, \emptyset, 0, 0]\) where the domain of~\(\dl\) is an empty set.

    \item \textbf{Introduce node.}
      Let~\(i\) be an introduce node with a child node~\(j\), then~\(X_i = X_j \cup \{u\}\) where~\(u \not\in X_j\).
      Let~\(D^\delta_j[\dl'_\delta, S', k_j, p_j]\) be an entry of the table of the child node~\(j\) with value~\(1\).
      We construct a~\(\delta\)\nobreakdash-labelling function~\(\dl_\delta\) which agrees with~\(\dl'_\delta\) on~\(X_j\) and tries all possible values for~\(u\).
      In particular the constructed~\(\delta\)\nobreakdash-labelling functions set~\(\dl_\delta(u)\) to values~\(\Sigma_\varrho \setminus \{0\}\) and additionally we try the label~\(0\) for~\(u\) if~\(u\) is a supplier.
      For such a~\(\delta\)\nobreakdash-labelling function~\(\dl_\delta\) we compute~\(S\) to be the set of satisfied vertices of~\(X_i\) as follows.
      We add the entire set~\(S'\) to~\(S\).
      We add~\(u\) to~\(S\) if there exists a neighbor~\(v \in N(u)\) so that~\(\dl_\delta(u) \geq \dl_\delta(v) + \frac{d(u, v)}{1 + \varepsilon}\).
      Finally, we add neighbors~\(w \in N(u)\) to~\(S\) for which it holds that~\(\dl_\delta(w) \geq \dl_\delta(u) + \frac{d(u, w)}{1 + \varepsilon}\).
      If~\(\dl_\delta(u) = 0\) and~\(k_j \leq k - 1\), then we set the entry~\(D^\delta_i[\dl_\delta, S, k_j + 1, p_j]\) to~\(1\).
      If~\(\dl_\delta(u) \in \Sigma_\varrho \setminus \{0, \infty\}\), then we set the entry~\(D^\delta_i[\dl_\delta, S, k_j, p_j]\) to~\(1\).
      If~\(\dl_\delta(u) = \infty\), \(u \in \vcust\), and~\(p_j \leq p - 1\), then we set the entry~\(D^\delta_i[\dl_\delta, S, k_j, p_j + 1]\) to~\(1\).
      If~\(\dl_\delta(u) = \infty\) and~\(u \in \vsup\), then we set the entry~\(D^\delta_i[\dl_\delta, S, k_j, p_j]\) to~\(1\).

    \item \textbf{Forget node.}
      Let~\(i\) be a forget node where~\(X_i = X_j \setminus \{u\}\) and~\(u \in X_j\).
      For any~\(\delta\)\nobreakdash-labelling function~\(\dl_\delta\) of~\(X_i\) we denote by~\(L_c(\dl_\delta)\) the set of functions which extend~\(\dl_\delta\) by assigning labels~\(\Sigma_\varrho \setminus \{0, \infty\}\) to~\(u\), by~\(L_s(\dl_\delta)\) the set of functions which extend~\(\dl_\delta\) by assigning labels~\(\Sigma_\varrho\) to~\(u\), and by~\(L_\infty(\dl_\delta)\) the extension of~\(\dl_\delta\) which assigns label~\(\infty\) to~\(u\).
      Let~\(S\) be any subset of clients of~\(X_i\), \(k_i \in \{0, \ldots, k\}\), and~\(p_i \in \{0, \ldots, p\}\).
      If~\(u\) is a supplier, then we set
      \begin{equation}
        D^\delta_i[\dl_\delta, S, k_i, p_i] = \bigvee_{\dl'_\delta \in L_s(\dl_\delta)} D^\delta_j[\dl'_\delta, S, k_i, p_i].
      \end{equation}
      If~\(u\) is a client, then we set
      \begin{equation}
        \begin{split}
          D^\delta_i[\dl_\delta, S, k_i, p_i] = & \left(\bigvee_{\dl'_\delta \in L_c(\dl_\delta)} D^\delta_j[\dl'_\delta, S \cup \{u\}, k_i, p_i]\right) \vee \\
                                                & \vee D^\delta_j[L_\infty(\dl_\delta), S, k_i, p_i].
        \end{split}
      \end{equation}

    \item \textbf{Join node.}
      Let~\(i\) be a join node with children~\(j_1\) and~\(j_2\) where~\(X_i = X_{j_1} = X_{j_2}\).
      Let~\(\dl_\delta\) be a~\(\delta\)\nobreakdash-labelling function on~\(X_i\).
      Then for each pair of true entries~\(D^\delta_{j_1}[\dl_\delta, S_1, k_1, p_1]\) and~\(D^\delta_{j_2}[\dl_\delta, S_2, k_2, p_2]\) we set to~\(1\) the entry
      \begin{equation}
        D_i\left[\dl_\delta, S_1 \cup S_2, k_1 + k_2 - \abs{\dl_\delta^{-1}(0)}, p_1 + p_2 - \abs{\dl_\delta^{-1}(\infty) \cap \vcust}\right].
      \end{equation}
  \end{itemize}

  To establish correctness of the algorithm, there are two tasks we need to accomplish.
  First, we need to show that for any bag~\(X_i\) we have~\(D^\delta_i[\dl_\delta, S, k_i, p_i] = 1\) if and only if there exists a~\(\delta\)\nobreakdash-labelling of~\(G[V_t]\) which agrees with~\(\dl_\delta\) on~\(X_i\), satisfies clients~\(((V_i \setminus X_i) \cup S) \cap \vcust\) aside from those clients with label~\(\infty\), and has cost~\(k_i\) and penalty~\(p_i\).
  The proof of this equivalence is done similarly to the proof of correctness of Theorem~\ref{thm:kswo-xp-alg-tw}.
  The only difference is that we need to consider~\(\delta\)\nobreakdash-labelling functions and~\(\varepsilon\)\nobreakdash-satisfiability instead of distance labelling and satisfiability respectively.
  Hence we omit this part of the proof.
  Using Lemma~\ref{lem:delta-labelling-equivalence}, we obtain a solution to the input \problemName{kSwO} instance of cost~\((1 + \varepsilon)^2\varrho\).

  It is more interesting to prove the following statement: we would like to show that if there exists a solution of cost~\(\varrho\), then there exists a~\(\delta\)\nobreakdash-labelling which is going to be found by the algorithm.
  The main difficulty of proving this statement is that the converse of Lemma~\ref{lem:delta-labelling-equivalence} does not hold for any choice of~\(\delta\).
  In the remainder suppose there exists a distance labelling~\(\dl \colon V \to \{0, \ldots, \varrho\} \cup \{\infty\}\) which encodes a solution to the instance as in the proof of Lemma~\ref{lem:kswo-epas-tw--dl}.

  Let~\(X_i\) be a bag of the decomposition of height~\(h\) and~\(S\) the vertices of~\(V_i\) satisfied by~\(\dl\) including suppliers.
  We are going to show that there always exists~\(\dl_\delta \colon X_i \to \Sigma_\varrho\), \(S_\delta \supseteq S\) and values~\(k_i\) and~\(p_i\) such that~\(D^\delta_i[\dl_\delta, S_\delta, k_i, p_i] = 1\), \(k_i \leq \abs{\dl^{-1}(0)}\), \(p_i \leq \abs{\dl^{-1}(\infty) \cap \vcust}\) and for all~\(u \in X_i\) we have~\(\dl_\delta(u) \in \left[\frac{\dl(u)}{(1 + \delta)^h}, (1 + \delta)^h\dl(u)\right]\).

  We prove this claim by induction on the height of a bag.
  This property trivially holds for empty leaf bags in the base case.
  For the induction step, consider a node at height~\(h + 1\).
  In the case of forget and join bags, if we assume that the desired property holds for their children, then it follows that it holds for them as well since~\(\dl_\delta(u) \in \left[\frac{\dl(u)}{(1 + \delta)^h}, (1 + \delta)^h\dl(u)\right]\) implies~\(\dl_\delta(u) \in \left[\frac{\dl(u)}{(1 + \delta)^{h + 1}}, (1 + \delta)^{h + 1}\dl(u)\right]\).

  It remains to prove the property for introduce nodes.
  Let~\(i\) be an introduce node with child~\(j\) where~\(X_i = X_j \cup \{u\}\), \(u \not\in X_j\).
  We cannot use the approach for proving the desired property we used for join and forget nodes.
  For the introduced vertex~\(u\) the induction hypothesis~\(\dl_\delta(u) \in \left[\frac{\dl(u)}{(1 + \delta)^h}, (1 + \delta)^h\dl(u)\right]\) does not apply when~\(\dl(u)\) is finite since~\(u \not\in V_j\).
  Let~\(S \subseteq X_i\) be the set of vertices (including suppliers) satisfied by~\(\dl\) in~\(V_i\) and similarly let~\(S' \subseteq X_j\) be the set of satisfied vertices in~\(V_j\).
  We claim that at least one of the following must be true:
  \begin{case}
    \label{kswo-epas-correctness-case-1}
    \(\dl(u) = 0\).
  \end{case}
  \begin{case}
    \label{kswo-epas-correctness-case-2}
    \(S = S' \cup \{u\}\).
  \end{case}
  \begin{case}
    \label{kswo-epas-correctness-case-3}
    \(u \not\in S\) and~\(\dl(u) < \infty\).
  \end{case}
  \begin{case}
    \label{kswo-epas-correctness-case-4}
    \(\dl(u) = \infty\).
  \end{case}
  Suppose for contradiction that~\(\dl(u) \in \{1, \ldots, \varrho\}\) and \(S \supseteq S' \cup \{u, v_1\}\) where~\(v_1 \in X_i \setminus S'\).
  If~\(v_1\) is satisfied in~\(X_i\) but not in~\(X_j\), then the sole cause of this fact is that~\(v_1\) is satisfied by~\(u\) as the labels of its neighbors aside from~\(u\) remain the same between~\(X_i \cap N(v_1)\) and~\(X_j \cap N(v_1)\).
  Thus we have~\(\dl(v_1) \geq \dl(u) + d(v_1, u)\).
  Since~\(u\) is a satisfied vertex and~\(\dl(u) \in \{1, \ldots, \varrho\}\), there exists a vertex~\(v_2 \in X_j\) such that~\(\dl(u) \geq \dl(v_2) + d(u, v_2)\).
  Together we have~\(\dl(v_1) \geq \dl(v_2) + d(v_2, u) + d(u, v_1) \geq \dl(v_2) + d(v_1, v_2)\) where the last inequality holds from the preprocessing using Lemma~\ref{lem:tree-decomposition-metric-closure-in-bag}.
  However, the last inequality shows that~\(v_1 \in S'\) which is a contradiction.

  We therefore need to establish that for each of the four cases above, the algorithm produces an entry~\(D^\delta_i[\dl_\delta, S_\delta, k_i, p_i]\) with~\(S \subseteq S_\delta\), \(k_i \leq \abs{\dl^{-1}(0) \cap V_i}\), \(p_i \leq \abs{\dl^{-1}(\infty) \cap V_i \cap \vcust}\), and~\(\dl_\delta(u)\) which is at most a factor~\((1 + \delta)^h\) apart from~\(\dl(u)\).
  Assume by the induction hypothesis that there exists an entry~\(D^\delta_j[\dl_\delta', S_\delta', k_j, p_j]\) with value~\(1\) for some~\(S_\delta' \supseteq S'\), \(k_j \leq \abs{\dl^{-1}(0) \cap V_j}\), \(p_j \leq \abs{\dl^{-1}(\infty) \cap V_j \cap \vcust}\) and~\(\dl_\delta'\) which has~\((\forall v \in X_j)\left(\dl_\delta'(v) \in \left[\frac{\dl(v)}{(1 + \delta)^{h - 1}}, (1 + \delta)^{h - 1}\dl(v)\right]\right)\).

  \begin{itemize}
    \item[Case~\ref{kswo-epas-correctness-case-1}]
      If~\(\dl(u) = 0\), the algorithm considers a~\(\delta\)\nobreakdash-labelling function~\(\dl_\delta\) which agrees with~\(\dl_\delta'\) on~\(X_j\) and sets~\(\dl_\delta(u) = 0\).
      Since the entry~\(D^\delta_j[\dl_\delta', S_\delta', k_j, p_j]\) has value~\(1\), the algorithm sets the entry~\(D^\delta_i[\dl_\delta, S_\delta, k_j + 1, p_j]\) to~\(1\) for some~\(S_\delta\).
      We claim that~\(S \subseteq S_\delta\).
      To see this, let~\(v \in S \setminus S'\).
      Then~\(v\) must be satisfied by~\(u\) and we have~\(\dl(v) \geq \dl(u) + d(u, v)\).
      From the induction hypothesis, \(\dl(u) = 0\), and~\((1 + \delta)^{h - 1} \leq 1 + \varepsilon\), we have~\(\dl_\delta(v) \geq \frac{\dl(v)}{(1 + \delta)^{h - 1}} \geq \frac{d(u, v)}{1 + \varepsilon}\).
      This shows that every vertex~\(v \in S \setminus S'\) is satisfied in~\(X_i\).
    \item[Case~\ref{kswo-epas-correctness-case-2}]
      Assume that~\(\dl(u) \not\in \{0, \infty\}\) and~\(u \in S\).
      Then there must exist a vertex~\(v\) which satisfies~\(u\), that is~\(\dl(u) \geq \dl(v) + d(u, v)\).
      Let~\(r = (1 + \delta)^{h - 1}\dl(u)\).
      The algorithm considers a \(\delta\)\nobreakdash-labelling function~\(\dl_\delta\) which agrees with~\(\dl_\delta'\) on~\(X_j\) and sets~\(\dl_\delta(u) = (1 + \delta)^{\lceil \log_{1 + \delta}r \rceil}\).
      Using~\((1 + \delta)^{h - 1} \leq 1 + \varepsilon\) we have~\(\dl_\delta(u) \geq (1 + \delta)^{h - 1}\dl(u) \geq (1 + \delta)^{h - 1}(\dl(v) + d(u, v)) \geq \dl_\delta(v) + \frac{d(u, v)}{1 + \varepsilon}\).
      Hence the algorithm correctly adds~\(u\) to~\(S_\delta'\) to obtain~\(S_\delta \supseteq S\).
      Moreover we have the required upper bound as well since
      \begin{equation}
        \dl_\delta(u) = (1 + \delta)^{\lceil \log_{1 + \delta}r \rceil} \leq (1 + \delta)^{\log_{1 + \delta}((1 + \delta)^{h - 1}\dl(u)) + 1} \leq (1 + \delta)^{h}\dl(u).
      \end{equation}
    \item[Case~\ref{kswo-epas-correctness-case-3}]
      Consider the case when~\(S \setminus S' \neq \emptyset \), otherwise there is nothing to prove.
      Let~\(v \in S \setminus S'\), then~\(v\) must be satisfied by~\(u\), that is~\(\dl(v) \geq \dl(u) + d(u, v)\).
      Let~\(r = \frac{\dl(u)}{(1 + \delta)^h}\).
      The algorithm considers a \(\delta\)\nobreakdash-labelling function~\(\dl_\delta\) which agrees with~\(\dl_\delta'\) on~\(X_j\) and sets~\(\dl_\delta(u) = (1 + \delta)^{\lceil \log_{1 + \delta}r \rceil}\).
      Using the induction hypothesis and~\((1 + \delta)^{h - 1} \leq 1 + \varepsilon\), we have
      \begin{equation}
        \dl_\delta(v) \geq \frac{\dl(v)}{(1 + \delta)^{h - 1}} \geq \frac{\dl(u)}{(1 + \delta)^{h - 1}} + \frac{d(u, v)}{1 + \varepsilon} \geq \dl_\delta(u) + \frac{d(u, v)}{1 + \varepsilon}.
      \end{equation}
      Hence the algorithm extends~\(S_\delta'\) by adding all elements of~\(S \setminus S'\) to create the set~\(S_\delta\).
      Moreover we have the required upper bound as well since
      \begin{equation}
        \dl_\delta(u) = (1 + \delta)^{\lceil \log_{1 + \delta}r \rceil} \leq (1 + \delta)^{\log_{1 + \delta}\frac{\dl(u)}{(1 + \delta)^h} + 1} \leq \frac{\dl(u)}{(1 + \delta)^{h - 1}}.
      \end{equation}
    \item[Case~\ref{kswo-epas-correctness-case-4}]
      The algorithm considers a \(\delta\)\nobreakdash-labelling function~\(\dl_\delta\) which agrees with~\(\dl_\delta'\) on~\(X_j\) and sets~\(\dl_\delta(u) = \infty\).
      Since a vertex with label~\(\infty\) cannot satisfy a neighbor by definition, we have~\(S_\delta = S_\delta'\).
      The desired bound on~\(\dl_\delta(u)\) for~\(u \in X_i\) follows trivially.
  \end{itemize}

  We conclude that whenever a feasible solution of cost~\(\varrho\) exists to the input instance, we are able to recover from the root bag of the dynamic programming table a solution of cost~\((1 + \varepsilon)^2\varrho\) with at most~\(k\)~centers and at most~\(p\)~outliers.
  In particular, there exists an entry in the dynamic programming table of the root bag~\(D_r[\dl_\delta, X_r \cap \vcust, k_r, p_r]\) where~\(\dl_\delta\) is a~\(\delta\)\nobreakdash-labelling of~\(X_r\) where for all~\(u \in X_r\) we have~\(\dl_\delta(u) \leq (1 + \delta)^H\dl(u) \leq (1 + \varepsilon)\varrho\), \(k_r \leq k\) and~\(p_r \leq p\).

  It remains to bound the running time of the algorithm.
  We have~\(\abs{\Sigma_\varrho} = \bigO(\log_{1 + \delta}\varrho) = \bigO\left(\frac{\log\varrho}{\log(1 + \delta)}\right) = \bigO\left(\frac{\log\varrho}{\delta}\right)\) where we use the fact that~\(\ln(1 + \delta) \approx \delta\) for sufficiently small~\(\delta\) (that is, sufficiently large~\(n\)).
  By setting~\(\delta = \Omega\left(\frac{\varepsilon}{\tw \cdot \log n}\right)\) and assuming~\(k, p \leq \abs{V(G)}\), we get a running time~\(\bigO^*\left((\tw \cdot \frac{\log n}{\varepsilon})^{\bigO(\tw)}\right)\).
  Using Lemma~\ref{lem:exp-in-log-n-is-fpt}, this is an FPT algorithm with running time~\(\bigO^*\left((\tw / \varepsilon)^{\bigO(\tw)}\right)\).
  \qed
\end{proof}

\section{Open Problems}

We conclude with the following open problems.
The algorithms given by Theorems~\ref{thm:ckswo-epas-dd} and~\ref{thm:kswo-epas-hd} have the number of outliers in the base of the exponent.
Is it possible to remove the outliers from the set of parameters?
An improvement of Theorem~\ref{thm:hd1-hardness-of-constant-approximation} would be to show that the hardness is preserved in the case of planar graphs.
It may be of interest that \problemName{Planar Capacitated Dominating Set} is~\(\classW{1}\)\nobreakdash-hard when parameterized by solution size~\cite{bodlaender2009planar}.
Goyal and Jaiswal~\cite{DBLP:journals/corr/abs-2110-14242} have shown that it is possible to 2\nobreakdash-approximate \problemName{CkC} when the parameter is only~\(k\), and that this result is tight.
An improvement of Theorem~\ref{thm:hd1-hardness-of-constant-approximation} would be to show that this lower bound is tight in low highway dimension graphs.
Finally, we ask whether there exists a problem which admits an EPAS in low highway dimension graphs but we cannot approximate in low doubling dimension graphs, i.e.~the converse of Theorems~\ref{thm:hd1-hardness-of-constant-approximation} and~\ref{thm:ckswo-epas-dd}.

\section*{Declarations}

\paragraph{Conflict of interest}
We acknowledge that all authors agreed to submit the manuscript without any conflict of interest.

\paragraph{Author Contribution declaration}
Tung Anh Vu carried out the research, developed the theoretical results, and drafted the manuscript.
Andreas Emil Feldmann provided supervision, guidance on the research direction, and contributed to the writing and revision of the manuscript.
Both authors read and approved the final manuscript.

\paragraph{Funding}
Andreas Emil Feldmann was supported by the project 19-27871X of GA ČR.
Tung Anh Vu was supported by projects 22-22997S and 24-10306S of GA ČR.

\bibliographystyle{splncs04}
\bibliography{ckc}

\appendix

\section*{List of symbols}

\begin{tabular}{|@{\hskip1em}l@{\hskip1em}|@{\hskip1em}p{\dimexpr\textwidth-2cm}|}
  \hline
  \(V_S\)                 & Set of suppliers in an instance of \problemName{CkS} and \problemName{CkSwO} \\
  \(V_C\)                 & Set of clients in an instance of \problemName{CkS} and \problemName{CkSwO}\\
  \(L\)                   & Capacity function of a \problemName{CkC} instance and its generalizations \\
  \(\phi\)                & Assignment function of a \problemName{CkC} instance and its generalizations \\
  \(B(u, r)\)             & Ball with center~\(u\) of radius~\(r\) \\
  \(\varrho\)                & Usually indicates the target cost of a solution of one of the problems in this work \\
  \(\hd\)                 & Highway dimension \\
  \(\gamma\)              & Universal constant in the definition of highway dimension (c.f.~Definition~\ref{defi:hd1}) \\
  \(\dd\)                 & Doubling dimension \\
  \(\langle m, n\rangle\) & Set of integers that are at least~\(m\) and at most~\(n\) \\
  \(\langle m\rangle\)    & \(\langle 1, m \rangle\) \\
  \(\tw\)                 & Treewidth \\
  \(X_t\)                 & A bag of a node~\(t\) of a tree decomposition \\
  \(\bot\)                & Value indicating that a client is an outlier \\
  \(\alpha\)              & Aspect ratio of a metric space \\
  \(\dl\)                 & Distance labelling function used in the algorithm for \problemName{kSwO} in Section~\ref{sec:kswo-epas-tw} \\
  \hline
\end{tabular}

\end{document}